\documentclass[useAMS,usenatbib,a4paper]{mn2e}
\usepackage{aas_macros}
\usepackage{graphicx}
\usepackage[usenames]{color}

\title[Spectroscopy of $z\sim5$ Galaxies]{Spectroscopy of $z\sim5$ Lyman Break Galaxies in the ESO Remote Galaxy Survey}

\author[L. S. Douglas et~al.]{L.S. Douglas$^{1,2}$, M.N. Bremer$^2$, M.D. Lehnert$^1$, E.R. Stanway$^1$ \& Bo Milvang-Jensen$^{3}$\\
$^1$Laboratoire d'Etudes des Galaxies, Etoiles, Physique et Instrumentation GEPI, Observatoire de Paris, Meudon, 92195 France\\
$^2$H H Wills Physics Laboratory, Tyndall Avenue, Bristol, BS8 1TL, UK\\
$^3$Dark Cosmology Centre, Niels Bohr Institute, University of Copenhagen, Juliane Maries Vej 30, 2100 Copenhagen, Denmark
}

\begin{document}

\date{Accepted . Received ; in original form }

\pagerange{\pageref{firstpage}--\pageref{lastpage}} \pubyear{}

\maketitle

\label{firstpage}

\begin{abstract}

 We present the global results of a large spectroscopic survey carried
 out in order to identify $z\sim 5$ Lyman break galaxies (LBGs) across
 ten widely-separated $\sim 45$ arcmin$^2$ fields to a depth of
 $I_{AB}=26.3$. The redshifts of seventy $4.6<z<5.6$ LBGs were
 identified through their Ly$\alpha$ emission and/or a strong
 continuum break, with thirty eight sources showing detectable line
 emission of between $2.6\times 10^{-18}$ and $7\times 10^{-17}$ erg
 cm$^{-2}$ s$^{-1}$. Just over half of the spectroscopically-confirmed
 $z\sim 5$ galaxies have rest-frame Ly$\alpha$ equivalent widths above
 20 \AA, double the frequency of similarly strong line emitters in
 similar $z\sim3$ LBG samples. However, when reasonable corrections
 are made for the spectroscopically-unconfirmed sources that are
 nevertheless at these redshifts in both samples,we find no
 significant difference in the frequency of high equivalent-width line
 emitters between the samples. The rest-frame UV continuum slope of a
 typical $z\sim5$ line-emitting galaxy (as measured primarily from
 photometry, but also apparent in spectroscopy) is bluer than that
 of a typical break-only galaxy, a difference that is difficult to
 explain purely by differences in the ages of their stellar
 populations. Variation in metallicity and/or dust extinction can more
 straightforwardly account for this difference. If the correlation
 between metallicity and UV continuum slope identified at low redshift
 is applicable at $z>3$, the typical $z\sim 5$ LBGs have metallicities
 a factor of three lower than those of LBGs at $z\sim3$. HST imaging
 of a subset of the LBGs indicates that a large majority of the
 spectroscopically-confirmed LBGs in our sample are members of
 multiple systems (on $\sim$ arcsec scales) and/or show disturbed
 morphology. Using local LBG analogues as a model, this multiplicity
 could be explained either by super-starburst regions within a larger
 unseen structure, or by a high incidence of merging events at this
 epoch. The current data cannot distinguish between these two
 possibilities.  The surface density of $z\sim5$ LBGs in two of the
 ten fields is considerably higher than in the rest.  Both show clear
 spikes in their redshift distributions indicating strong
 three-dimensional clustering in these fields. Against an expectation
 of about one source per 0.1 in redshift between $4.8<z<5.6$, one
 field has seven identified objects between $5.11<z<5.21$ and the
 other has 17 between $4.95<z<5.15$. Neither structure can be bound
 given their depth in redshift and probably extend beyond the observed
 fields. The three-dimensional distances between LBGs in the structures
 are too large for them to have triggered their starbursts through
 mutual gravitational interaction, and so it is likely that the short-lived
 LBGs represent only a small fraction of the baryons in the
 structures.

\end{abstract}

\begin{keywords}
galaxies: high-redshift, galaxies: starburst
\end{keywords}

\section{Introduction}
\label{sec:intro}

The identification and observational study of the earliest galaxies
is crucial to our understanding of how galaxies form and evolve. With
increasing cosmic age, increasingly complex physical processes
can affect the properties of galaxies. Consequently, it is
hoped that the earlier a galaxy is observed, the more straightforward
it will be to compare its properties to any theory of galaxy
growth. Whether or not this turns out to be true in practice, such
galaxies can be used as markers of early structure
formation and explored as candidate emitters of the radiation
responsible for reionization.

While several different techniques have been successfully used to
identify galaxies at $z>5$, the Lyman break technique is arguably the
most useful as it can provide a sample of objects with measurable properties for individual galaxies (Lyman
$\alpha$ line strength, continuum spectral energy distribution and
potentially absorption line diagnostics of the sample or of individual
objects if bright enough).  Optically-selected samples of high redshift
Lyman break galaxies (LBGs) are dominated by systems that are undergoing strong and
unobscured star formation because by $z\sim 5$ optical
observations probe the rest-frame spectral region shortward of
1800\AA, where emission is dominated by comparatively short-lived O
and B stars.

In this decade there has been considerable progress
in the observational identification of systems at these
redshifts. This has allowed us to take the first  steps towards
understanding the properties of galaxies that most likely represent
the earliest stages in the evolution of current-era massive galaxies
\citep[\textit{ e.g.}][]{verma07}. Thousands of candidate $z\sim 5$ star
forming LBGs have been identified in large-area
ground-based photometric surveys, but only a very small fraction of
these have been confirmed spectroscopically. While studies of
optically-selected photometric samples can provide reliable insights
into the properties of the sources when HST imaging is used on its own
or in combination with deep infra-red photometry
\citep[\textit{e.g.}][]{bouwens07,verma07}, deriving reliable conclusions from
ground-based optically-selected samples is more problematic. Not only
do these samples suffer from the effects of contamination \citep[see
{\it e.g.}][hereafter paper I]{stanway08b, douglas09}, but the information supplied
on each system is minimal and so their detailed properties cannot be
explored.

Spectroscopically-confirmed samples of $z\sim5$, while inevitably
smaller than photometric samples, ultimately allow more detailed studies of
the properties of the galaxies. With a reliable sample, uncertainties
in statistical properties such as the luminosity function can be
minimised. Correlations between observables, such as line-strength
and colour, start to unravel the evolution of the objects. Instead
of exploring the clustering by statistical methods in two dimensions (which can be
unreliable if the sample is contaminated), a direct determination of
three-dimensional clustering can be made. With the stochastic and
relatively short-lived nature of the star formation within these
systems, the interpretation of clustering is potentially complicated
given that the timescale over which a source may be UV-luminous can be
only a small fraction of the look-back time through a survey
volume. The most important benefit of a spectroscopically-confirmed
sample is that it allows further spectroscopic follow-up in other
wavebands \citep[\textit{e.g.}][]{stanway08a}, so will eventually lead to a
fuller understanding of the early stages of galaxy formation and
evolution.

In order to furnish ourselves with a sample of objects with which to
carry out such studies, we have carried out a spectroscopic survey of
ten widely-separated $\sim 45$ arcmin$^2$ fields as part of the ESO
Remote Galaxy Survey (ERGS), an ESO large programme (PI M. Bremer, ID
175.A-0706) in order to identify and spectroscopically-confirm a
sample of $z\sim 5$ unobscured star-forming galaxies. We had several
motivations for carrying out this spectroscopic survey. Firstly, to
explore the reliability of photometric samples and how they impact the
reliability of statistical properties for $z\sim 5$ LBGs,  {\it e.g.} their
luminosity function (see paper I). Secondly, to obtain as large a sample of
confirmed $z\sim 5$ LBGs as possible in a reasonable amount of time
(the aim being at least 50 $z\sim 5$ LBGs in 100 hours of 8m
spectroscopy). A sample of this size (especially when combined with
other published samples) allows exploration of the properties of the
sources and of any correlation between them. Thirdly, by obtaining
such a sample over multiple widely-separated pointings, 3-dimensional
clustering can be explored directly, without being dominated by
cosmic variance in any one field. 

The structure of the paper is as follows. In section
\ref{sec:observations} we describe the data and sample selection of
high redshift LBGs. Details and examples of spectroscopic
identification are given in section \ref{sec:specres}. Discussion of
the global properties of the sample follows in subsequent
sections. The $z\sim5$ redshift distribution is described in section
\ref{sec:reddist}. Section \ref{sec:colours} explores the rest-frame
UV continuum of the LBGs and in section \ref{sec:ew} the properties of
the observed Ly$\alpha$ emission are discussed. In section
\ref{sec:specmorph} we describe the morphology of the confirmed high
redshift LBGs.  The surface density of sources is discussed in section
\ref{sec:surface} and the results of the $z\sim6$ LBG search are
presented in section \ref{sec:z6}.

We adopt the standard $\Lambda$CDM cosmology, i.e. a flat universe
with $\Omega_{\Lambda}=0.7$, $\Omega_{M}=0.3$ and $H_{0}=70
{\rm km\,s}^{-1}\,{\rm Mpc}^{-1}$.  All photometry was determined in
the AB magnitude system \citep{oke83}.

\section{Data and Sample Selection}
\label{sec:observations}

\subsection{Imaging Data}

The spectroscopic sample was derived from the same set of imaging data
used to derive the photometric sample previously presented in paper
I. The data included deep (around two hours per band)
$V$, $R$, $I$ and $z$-band imaging obtained with FORS2 \citep{appen98}
on the VLT and moderately-deep $J$ and $K_{s}$-band imaging 
(5-6 hours per band) obtained with SOFI on the NTT.  The optical data
covered approximately 49 arcmin$^2$ in each field, while the near-IR
data covered approximately half the area of the optical data per
field. While the photometric sample reported in paper I was drawn from
286 arcmin$^2$ requiring both optical and near-IR coverage, the
spectroscopic candidates were drawn from an effective area of 440
arcmin$^2$, the area covered by the $R$ and $I$-band imaging
after taking into account the area compromised by high surface
brightness foreground objects. All
imaging except the $z$-band frame were obtained as part of the
ESO Distant Cluster Survey (EDisCS), itself an ESO large programme \citep[PI
S. White, ID:166.A-0162, see][]{white05}. The $z_{AB}$-band
imaging was obtained as part of ERGS. Although there was some
field-to-field variation, typical $2\sigma$ limits in a 2$''$ circular
aperture were $V_{AB}$=28.1, $R_{AB}$=28.0, $I_{AB}$=27.1,
$z_{AB}$=26.0, $J_{AB}$=24.5, $K_{AB}$=23.6 and seeing varied between
$\sim 0.6''$ and $0.8''$ in the optical bands, $0.8''$ and $1.0''$ in
$J_{AB}$ and $0.6''$ and $0.8''$ in $K_{AB}$, comparable to the quality and depth
of other surveys of the equivalent or larger area. See paper I for a detailed 
field-to-field breakdown.

\subsection {Spectroscopic Sample}
\label{sec:sample}

\begin{figure}
\begin{center}
\includegraphics[width=8cm]{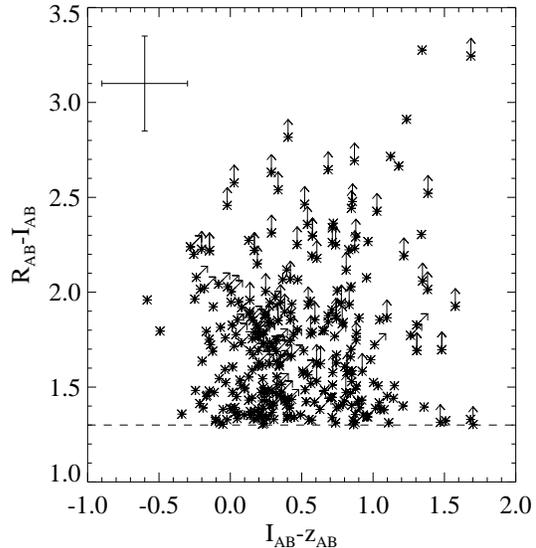}
\caption{Colour-colour diagram of the spectroscopically observed objects illustrating the range of colours sampled. The dashed line is the formal colour cut used in the photometric selection. Arrows depict where 2$\sigma$ limits were used.}
\label{fig:targets}
\end{center}
\end{figure}

As for the photometric sample discussed in paper I, the primary
selection criteria of $I_{AB}<26.3$, $R_{AB}-I_{AB}>1.3$ and
$V_{AB}<27.5$ were used to select objects that may have the
characteristic spectral break of $z>4.8$ LBGs. The same SExtractor
\citep{bertin96} catalogues were used to select objects for the
photometric sample and spectroscopic targets. The photometric sample
was generated using 2 arcsec diameter circular apertures so that
photometry across all the bands (with different pixel sizes) could be
combined in the object selection. While the photometric sample was
defined to allow an accurate determination of completeness and
contamination, one aim of the spectroscopy was to maximise the number
of spectroscopic candidates to be targeted with the slit masks,
thereby maximising the size of the spectroscopically-confirmed
$z\sim5$ LBG sample. Consequently, we performed the candidate selection twice
using the same criteria with different photometric catalogues. The first catalogue 
used the aperture magnitudes from paper I and the second used total magnitudes
measured using the SExtractor MAG\_AUTO apertures defined in $I$ and with appropriate corrections to the
$V$ and $R$-band value (typically of a few hundredths of a
magnitude due to the difference in seeing between the bands). If an
object satisfied the primary selection criteria in either of the apertures,
it was flagged as a candidate for spectroscopy. In addition, as we
specifically wanted to both maximise the number of candidates placed
on the spectroscopic masks and explore the potentially
contaminating populations, we avoided an explicit $I_{AB}-z_{AB}$ colour cut to
rule out potential lower redshift interlopers. In designing the
observational strategy, we matched the number of masks to be used in
any field to the number of objects selected by the primary
selection criteria, ensuring that there were generally sufficient slits to
observe objects with any $I_{AB}-z_{AB}$ colour.  Figure \ref{fig:targets} shows
the range of colours of the objects observed during the spectroscopy
program.

Even with several masks per field, pairs of
candidates will compete for space because their
positions are such that their dispersed spectra would
overlap. Consequently the candidates were prioritised drawing on our
previous experience of comparable spectroscopy with the same
instrumentation. We had previously observed four adjoining FORS2
fields, searching for $z\sim5$ LBGs \citep[][Davies et al. in
prep]{lehnert03}. As a result of this experience we were able to
prioritise the candidates based on their likelihood of being at high
redshift, as described below. This was done to maximise the number confirmed of $z\sim 5$ LBGs on the masks.

Each candidate was categorised according to its photometry in all
bands and if necessary from visual inspection of all of the
ground-based images. If the object had $I_{AB}-z_{AB}<1$, was
undetected in the $V$, $J$ and $K_{s}$-band images, and the photometry was
in no way influenced by neighbouring objects (affecting the background
estimate or the positioning of the aperture), the object was believed to
be at high redshift and given the highest priority (noted as priority 1 in
Tables 1 and 2). If the object was detected in the $V$-band,
indicating that it may be a lower redshift galaxy or a star, or had
potentially compromised photometry due to object overcrowding it was
assigned a lower priority (priority 2 in Tables 1 and 2). Sources
which were detected in the near-infrared imaging, with colours
inconsistent with high redshift star-forming galaxies were classified
as likely `low z' objects. Finally, in empty areas of the
spectroscopic masks where no other strong candidates could be placed,
objects with photometry outside the selection criteria ({\it e.g.}
objects with red $R_{AB}-I_{AB}$ colours but $I_{AB}>$26.3, objects with
$1.0<R_{AB}-I_{AB}<1.3$, or objects detected in the $V$-band) were observed,
classified as `other' sources. This was particularly valuable as it
probed additional colour space outside the selection. In practice,
sufficient objects  were observed to obtain good statistics for
each class despite slit contention.

Our previous experience had shown that brighter ($I_{AB}<25$)
candidates were almost invariably low redshift galaxies ($0.6<z<1.2$)
or stars, especially if they had red $I_{AB}-z_{AB}$ or
$I_{AB}-K_{AB}$ colours (see paper I for details of the colour
modelling of the contaminating populations). However, even in the
cases where the multi-band characteristics of an object strongly
indicated a low or zero-redshift, an object was not rejected from the
list of candidates, merely given a lower priority. This was done so
that the reliability of our selection, both for this sample and also
for the photometric sample in paper I, could be determined. For
objects with $I_{AB}<25$, it is impossible to rule out with 100 per
cent certainty an object being at high redshift based on the
ground-based photometry alone; all that can be determined is that an
object is highly unlikely to be at high redshift.  Objects brighter
than $I_{AB}<24$ selected by our colour cuts would most likely have a
strong AGN component in order to be at $z\sim 5$. While our optical colours do
not distinguish between a $z\sim 5$ quasar and certain types of
late-type stars, given that we detected only a single fainter object
with a moderate AGN component in our sample \citep[see][]{douglas07},
it is likely that no brighter AGN existed in the volume probed by the
survey.

As the near-IR data did not cover all of the optical fields, the full
prioritisation could not be carried out uniformly across all
areas. This resulted in less rigorous prioritisation in a fraction of
each field. Fortunately, objects which were near-IR detected and
therefore less likely to be at the highest redshifts tended to be the
brighter objects in the optical, or had red $I_{AB}-z_{AB}$ colours,
both of which factors downgraded their priority on their own.  Thus a
lack of near-IR data did not dramatically change the prioritisation of
objects between the regions with and without near-IR data. As
prioritisation was only important when there was contention between
candidates for space on the slit masks, many candidates were observed
regardless of their priority. Table \ref{masks} shows the fraction of
high and low redshift targets which were observed. The results in
table \ref{tab:expected} (discussed below) confirm {\it a posteriori}
that the prioritisation was effective.

\begin{table*}
\caption{The number of targets which were observed spectroscopically in the full optical field of view. Targets include high redshift candidates divided into priority 1 and 2 (see text) and likely lower redshift candidates which obey the optical selection criteria but are detected in the near-IR imaging. For each field, the number of photometrically selected targets and the number observed on mask are given for each target category.}
\begin{center}
\begin{tabular}{|c|c|c|c|c|c|c|c|c|c|c|}
\hline
Field &  No. & \multicolumn{2}{|c|}{Priority 1 $z\sim5$} &  \multicolumn{2}{|c|}{Priority 2 $z\sim5$} & \multicolumn{2}{|c|}{Low z targets}\\
&  masks &  targets & on mask & targets & on mask &  targets & on mask   \\ \hline
J1037.9-1243 & 1 & 7 & 5 & 8 & 2 & 5 & 3 \\
J1040.7-1155 & 3 & 30 & 24 & 12 & 4 & 12 & 7 \\
J1054.4-1146 & 1 & 7 & 4 & 6 & 3 & 5 & 2 \\
J1054.7-1245 & 5 & 53 & 48 & 27 & 15 & 11 & 4 \\
J1103.7-1245 & 2 & 7 & 5 & 8 & 7 & 3 & 3\\
J1122.9-1136 & 1 & 8 & 3 & 10 & 4 & 13 & 4 \\
J1138.2-1133 & 1 & 7 & 6 & 8 & 3 & 5 & 0 \\
J1216.8-1201 & 2 & 32 & 9 & 23 & 9 & 11 & 9 \\
J1227.9-1138 & 2 & 14 & 13 & 28 & 11 & 5 & 5 \\
J1354.2-1230 & 2 & 15 & 12 & 12 & 4 & 11 & 9 \\ \hline
All      &     & 180 & 129 (72\%) & 132 & 62 (47\%) & 81 & 46 (57\%) \\ \hline
\end{tabular}
\end{center}
\label{masks}
\end{table*}

In addition to the $z\sim 5$ target selection, our data allowed us to
select $z\sim6$ candidates using a redder $I_{AB}-z_{AB}>1.3$
colour cut which at $z_{AB}>24.5$ will select LBGs at $z>6$
\citep[\textit{e.g.}][]{stanway04}.  Such objects will not be detected in our
near-IR imaging, or in the $V$-band image. These higher redshift LBG candidates were
rarer than those at $z\sim 5$. There were typically nine such
candidates per field and a total of 73 candidates were observed on the
masks, placed in regions where they did not compete for space with the
highest priority $z\sim 5$ candidates.

\subsection{Observations}

Twenty VLT/FORS2 `MXU' spectroscopic masks were used to obtain redshifts in
the ten fields. The number of masks assigned to each field broadly
reflected the number of high priority $z\sim5$ candidates. The only
exception is that the field of J1103.7 received an extra mask at the
expense of field J1216.8 because the placement of candidates in the
latter field made for an inefficient third mask. Each mask was
designed to have between 30 to 40 slitlets of width one arcsecond and
typical length of 10\,arcsec. In general, where two or more candidates
contended for space on a slit mask, the highest priority object was
chosen. Occasionally, in order to help calibrate the reliability of
the photometric sample, a potential contaminant was chosen at random
over a higher priority object. For this reason, and because high
priority candidates will inevitably contend with each other for space
on the mask, not all of the high priority targets had slitlets
assigned to them (see table \ref{masks}).

The observations with FORS2 were carried out in service mode on
nights between December 2005 and May 2006. Seeing was always
better than 1\,arcsec and observations
were taken during dark time, clear or photometric conditions and
at an airmass below two.  The masks were made using {\it fims}, the
mask design software for FORS2 distributed by ESO. Each mask was
observed in five blocks of 43\,minutes. Each block consisted of four
observations of 650\,seconds at four different nod positions along the
slit resulting in a total on source exposure time of 3.6\,hours. These nod
positions were also offset between blocks ensuring that the spectrum
of each object was placed at 20 independent positions on the CCD. This
technique was used to avoid the effects of pattern noise or bad pixels
and to facilitate the best possible sky subtraction. FORS2 was used in
spectroscopic mode using the 300I grism with the OG525 blocking
filter to restrict the wavelength range to between 650-1000nm. The
resulting pixel scale was 3.2\AA\ in the spectral direction by 0.25\,arcsec 
spatially and the spectral resolution was $\sim10$\AA, measured
from the width of the sky lines. This observing setup was identical to
that used by \citet{lehnert03} and the choice of grism and blocking
filter the same as that used by \citet{vanzella05} and
\citet{vanzella09}.

\subsection{Data Reduction}
\label{sec:dreduct}

The data were reduced following the same
standard procedures as those described in \citet{lehnert03}, with
the following modifications. The individual spectra were rectified
 before sky subtraction. As the seeing of the
spectroscopic observations was close to (but always less than) the
slit width, the spectroscopy did not recover all of the flux of an
object. The standard star spectra were taken through a 5 arcsec-wide
slit and so recovered a higher fraction of the stellar flux. To account for
this, the ratio of flux recovered in 1$''$ and 5$''$ wide rectangular apertures
was measured from $I_{AB}$-band imaging of standard stars, and the fluxes
determined in the spectroscopy were corrected by the factor
appropriate to the seeing of the individual exposures. The correction factors
typically ranged from 1.1 to 1.5. The final combined exposures reached
a flux limit of a few 10$^{-18}$\,ergs\,cm$^{-2}$\,s$^{-1}$\,\AA$^{-1}$ for
an unresolved emission line in a dark region of the sky spectrum,
comparable to the depth obtained by \citet{lehnert03} as expected given
the identical observational setup (exposure time, grism, slit width).

\section{Spectroscopic Identifications}
\label{sec:specres}

The reduced spectra were examined by four of the authors (LSD, MNB,
ERS and MDL) and classified
into high redshift, low or intermediate redshift and low
signal-to-noise categories. The high redshift sources were expected to
show a clear continuum break at the redshifted wavelength of
Ly$\alpha$, often with a Ly$\alpha$ emission line at the same
wavelength. Low or intermediate redshift objects were identified
through multiple emission lines (such as H$\beta$ and [OIII]), no
break in the continuum or a steeply reddening continuum with molecular
absorption bands indicative of low mass stars \citep{stanway08c}. Low
redshift objects were also identified by the presence of continuum
emission at the bluest wavelengths ($\sim$6500\AA),
inconsistent with a red $R_{AB}-I_{AB}$ colour and therefore included in the
sample due to photometric uncertainty.  Any spectra exhibiting one or
more of these features were classified as low redshift. The object did
not necessarily have to have a secure redshift, merely that the
spectrum was clearly inconsistent with that of a high redshift
LBG. Objects where the signal-to-noise in the continuum was too low to
be placed in either of the above categories were classified as
`unknown'.

In addition to categorising an object as being at high redshift
($z\sim 5$), we split this category into two grades: A and B. If the
spectrum was clearly that of a high redshift object with an emission
line at the wavelength of a continuum break, or a clear continuum
break with no other obvious low redshift features red-ward or
blue-ward of the break, the spectrum was classified as grade A. The
redshift was determined to the resolution limit of the spectrograph,
$\sim$400 km\,s$^{-1}$, depending upon the location of the features. If
the continuum break or Ly$\alpha$ emission line was comparatively
weak, particularly if these features were located in the spectral
regions containing strong skyline residuals limiting the accuracy of
the redshift determination to $\sim1000$ km\,s$^{-1}$, it was
classified as grade B. We visually compared our spectra to the
published one and two-dimensional spectra of \citet{vanzella09} and
found that the quality of our grade A and B spectra are broadly
similar to that of similarly classified spectra in their work, even
though the details of the classification schemes are not identical.

Over the ten fields, 54 sources were classified as grade A $z>4.6$
galaxies, 36 of which were clear Ly$\alpha$ line emitters and 18
showed only a Lyman break at the signal-to-noise of our data. A
further 16 candidates at $z>4.6$, but with grade B spectra were
confirmed, two Ly$\alpha$ emitters with weak lines and 14 continuum
break galaxies. The redshifts of most of these placed Ly$\alpha$ in
spectral regions of comparatively high sky noise caused by the strong
atmospheric lines.  Table \ref{tab:expected} lists the number of high
and low redshift galaxies which were confirmed within the different
candidate sub-samples.

\begin{table}
\caption{Number of galaxies in the full optical field of view confirmed at low or high redshift according to their assumed low or high redshift photometric identification. Rows give the candidate sub-sample designation, columns give the resulting spectroscopic outcome. The percentages in brackets relate to the fraction of the candidate sub-sample resulting in either a high redshift or unknown classification. The low redshift objects are split into those drawn from areas with and without nIR data. The different sub-samples are described in section \ref{sec:sample}}
\begin{center}
\begin{tabular}{|c|c|c|c|c|}
\hline
Sample &   $z\sim5$ & \multicolumn{2}{|c|}{Low z objects} & Unknown\\ 
 &    & with nIR data & no nIR data& (Low S/N)\\ \hline
Pri 1  & 54 (42\%) & 8 & 10 & 57 (44\%)\\
Pri 2  & 8 (13\%) & 8 & 8 & 38 (61\%)\\ 
low z & 2  (4\%)& 30 & - & 14 (30\%)\\ 
other & 6 (5\%)& 45 & 17 & 45 (40\%)\\ \hline
\end{tabular}
\end{center}
\label{tab:expected}
\end{table}

As can be seen in table \ref{tab:expected}, the vast majority of
galaxies confirmed to be at $z\sim5$ were drawn from the highest
priority sample.  However, a small number of galaxies
which had originally been flagged as likely low redshift interlopers
were found to be at high redshift. Similarly some objects flagged as
likely high redshift galaxies were confirmed to be at low redshift,
highlighting the limitations of purely photometric surveys. An
accurate redshift for some of the confirmed low redshift objects could
not be measured due to low signal-to-noise or featureless continuum,
however a $z\sim5$ LBG identification could be ruled out, usually
because of detectable continuum blue-ward of the expected Lyman break
at $z\sim 4.6$. 

A further six LBGs were identified in the `other' sample with photometry
outside of the selection criteria.  Three of these have a
$V$-band flux 0.1-0.2 magnitudes brighter than the $V$-limit (but
fulfil the $I$ and $R-I$ selection).  Two were fainter than the
$I$-band flux limit by 0.2 and 0.3 magnitudes.  The sixth galaxy
had a $R_{AB}-I_{AB}=1.03$, bluer than the selection limit. This can be
explained by the very strong Ly$\alpha$ emission line detected at
$z=4.9$ from this object. At this redshift the line emission falls within
the boundaries of the $R$-band filter, reducing the observed flux
difference between the $R$ and $I$-bands
\citep[see][]{stanway08b}.

\subsection{Example spectra}

\begin{figure}
\begin{center}
\includegraphics[width=8cm]{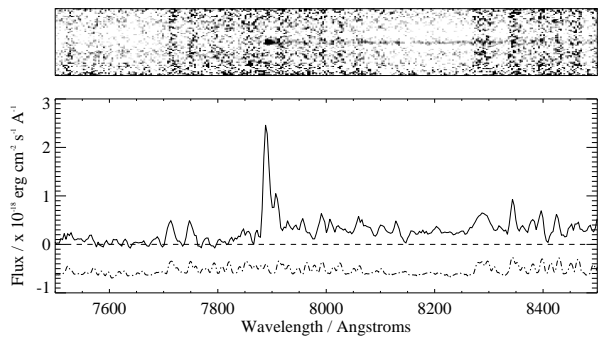}
\caption{Example of a high redshift galaxy at $z=5.5$. The top panel
 shows the reduced two dimensional spectrum where the horizontal
 direction is wavelength and the vertical direction is the spatial
 distance along the slit. The bottom panel shows a 1D extraction
 from the flux calibrated spectrum and the typical atmospheric emission
 at these wavelengths.}
\label{fig:highzex}
\end{center}
\end{figure}

Figure \ref{fig:highzex} shows an example of a $z=5.5$ galaxy. 
At the centre of the 2D spectrum there is clear
evidence of continuum emission with a break across an emission line, a
signature of Lyman $\alpha$ at high redshift. The continuum break and
emission line can also be clearly seen in the one-dimensional
spectrum. The relatively noisy vertical features in the 2D spectrum are residuals
from the subtraction of OH atmospheric sky emission. Small-scale
apparent continuum features seen in the spectrum can all be
accounted for by the sky lines, atmospheric
absorption or noise. 

\begin{figure}
\begin{center}
\includegraphics[width=8cm]{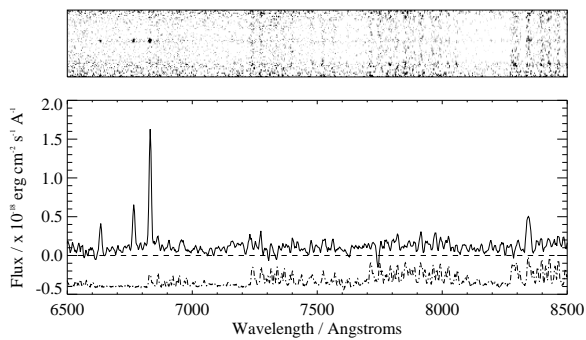}
\caption{Example of a low z emission line galaxy with H$\beta$ and the
 [OIII]$\lambda\lambda$4959,5007 doublet at $z=0.36$. Format as described in figure \ref{fig:highzex}.}
\label{fig:lowzex}
\end{center}
\end{figure}

\begin{figure}
\begin{center}
\includegraphics[width=8cm]{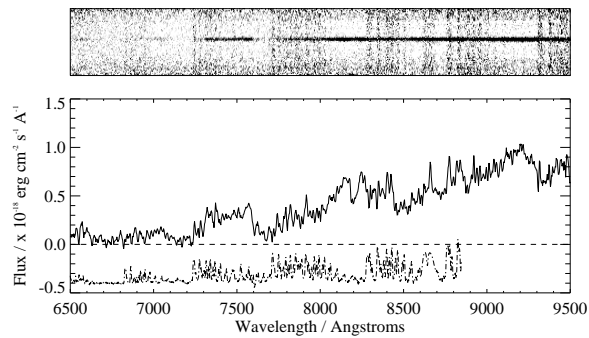}
\caption{Example of an M-type dwarf star \citep[also see][]{stanway08c}. Format as described in figure \ref{fig:highzex}. }
\label{fig:starex}
\end{center}
\end{figure}

Figures \ref{fig:lowzex} and \ref{fig:starex} show examples of low
redshift contaminants: an intermediate redshift emission line galaxy and a cool
star. Several strong emission lines are seen in figure
\ref{fig:lowzex} as well as detected continuum at the blue end of the 
spectrum. The line emission was identified as H$\beta\lambda$4861, and 
the [OIII]$\lambda\lambda$4959,5007 doublet placing the galaxy at a
redshift of $z=0.36$.

The stellar spectrum in figure \ref{fig:starex} shows a gradual
increase of continuum into the red, no break, and strong molecular
absorption bands at 7700\AA, 8500\AA\ and 9300\AA. At fainter
magnitudes spectra like this may look like a continuum source with a
strong break so extra care must be taken when identifying Lyman break
galaxies without line emission. It is more likely that they will appear
as a noisy continuum source which is comparatively strong at the red
end of the spectral range, gradually becoming fainter with decreasing
wavelength. In this case the source would be spectroscopically classified as 'unknown'

\begin{figure}
\begin{center}
\includegraphics[width=8cm]{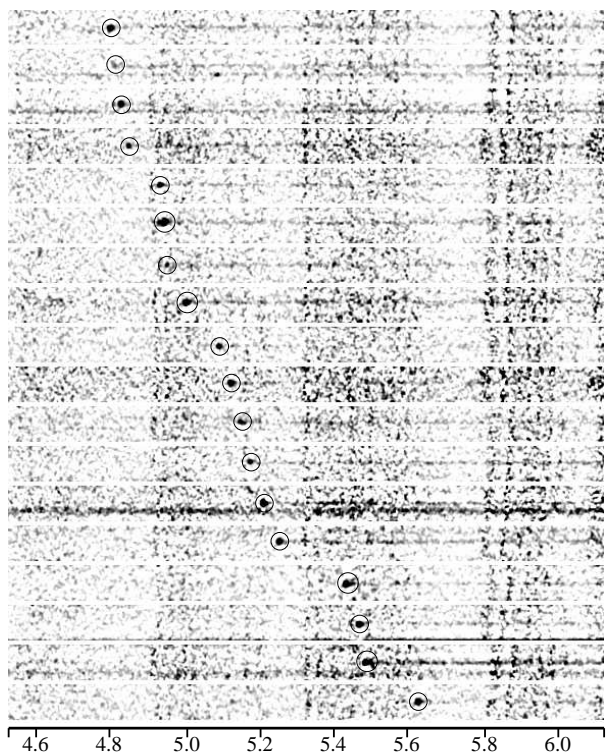}
\caption{Two dimensional spectra of a selection of Ly$\alpha$ emitters at increasing redshift with the Ly$\alpha$ emission marked by the circles. The vertical noise features caused by sky line residuals are clearly seen but do not appear to significantly effect the detection of emission lines. As these spectra have not been rectified for individual spectral distortions, the lower Ly$\alpha$ redshift scale is approximate.}
\label{fig:highzlines}
\end{center}
\end{figure}

\begin{figure}
\begin{center}
\includegraphics[width=8cm]{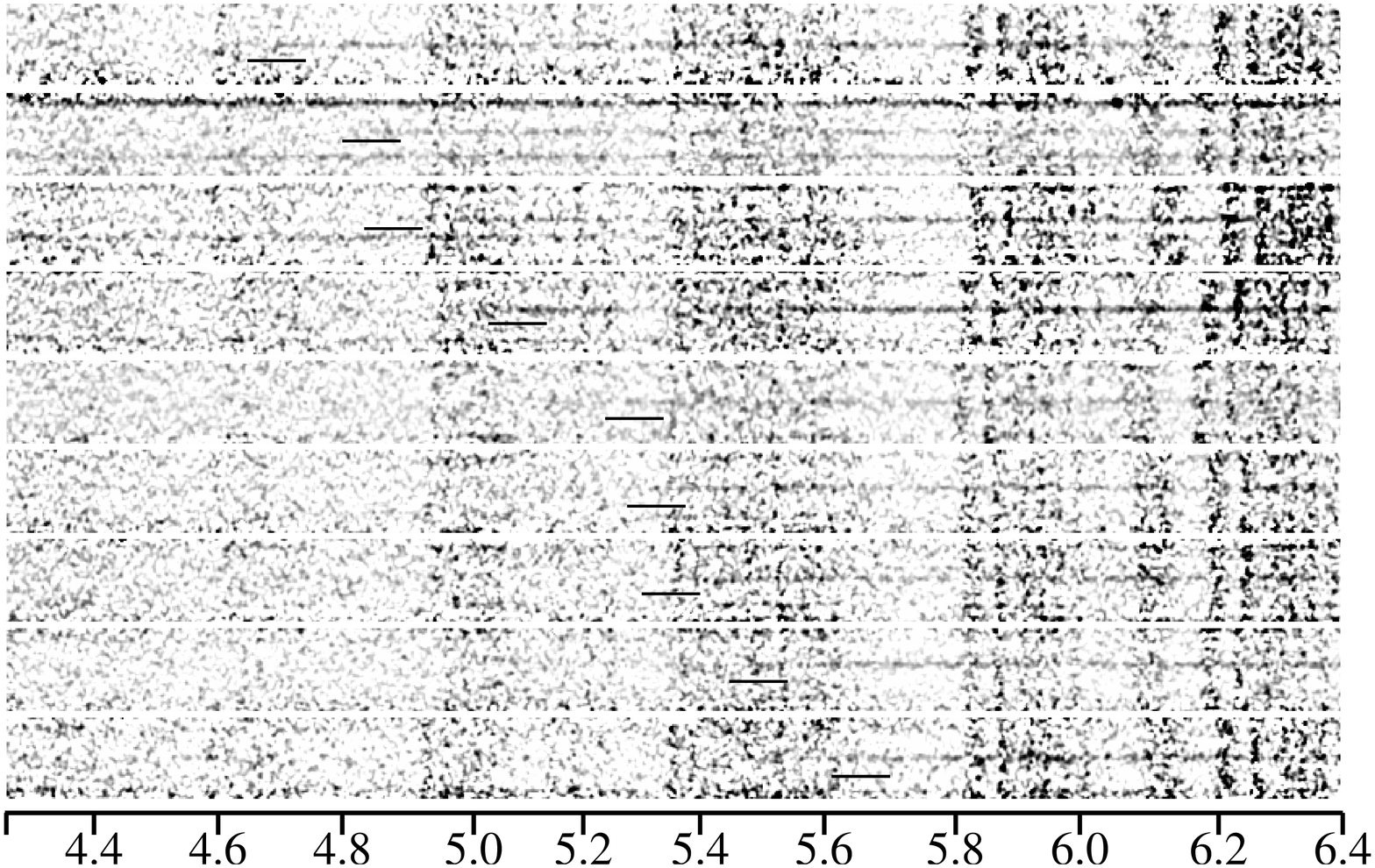}
\caption{Two dimensional spectra of a selection of Lyman break galaxies at increasing redshift with the location of the break marked by the black lines. Given
the faint nature of the objects it can be difficult to clearly display the continuum emission in this format. The vertical noise features are caused by sky line residuals. As the spectra have not been rectified for individual spectral distortions, the lower Ly$\alpha$ redshift scale is approximate.}
\label{fig:highzbreaks}
\end{center}
\end{figure}

Further examples of two-dimensional spectra of
line emitting and break-only galaxies can be seen in figures
\ref{fig:highzlines} and \ref{fig:highzbreaks}.

\section{Redshift distribution of confirmed $z\sim5$ galaxies}
\label{sec:reddist}

\begin{figure}
\begin{center}
\includegraphics[width=8cm]{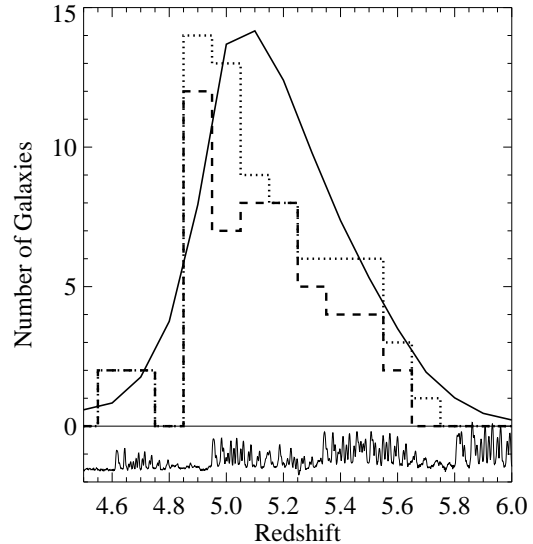}
\caption{Redshift distribution of spectroscopically confirmed galaxies. The dotted line represents all galaxies with redshifts $z>4.5$, the dashed line shows only grade A galaxies with secure redshifts and the solid line illustrates the shape of the predicted redshift distribution (scaled to match the peak number count of the observed distribution). Also included, below the distributions, is the atmospheric emission spectrum at the redshift of Ly$\alpha$.}
\label{fig:redist}
\end{center}
\end{figure}

The redshift distribution of all the spectroscopically-confirmed $z\sim 5$
ERGS sources is shown in figure \ref{fig:redist}. The solid
line represents the predicted redshift distribution of the candidates
given the photometric selection criteria, depth of images and the luminosity 
function of $z\sim5$ LBGs as determined in paper I. The
observed redshift distributions are shown by the dotted and dashed
lines. The grade A galaxies are shown by the dashed line and the whole
sample of grade A and B objects is shown by the dotted line. The
general shape of the observed distributions agrees well with the
prediction, indicating that the parameters of the simulations
(intrinsic colours of the galaxies, shape of the luminosity function
{\it etc.}) were reasonable. The atmospheric emission, dominated by
OH emission lines, at the wavelengths of Ly$\alpha$ for each redshift is
plotted below the redshift distributions. Identifying faint high
redshift galaxies, especially from breaks alone, in regions of high
noise caused by sky line residuals, can be very difficult. For the
strong Ly$\alpha$ line emission and continuum break galaxies (grade A
objects) this noise does not have a significant effect, but for
fainter sources areas of high noise can seriously hinder any
spectroscopic confirmation. This is reflected in figure
\ref{fig:redist} where the majority of objects with grade B
identifications are in areas of high sky signal (and hence large
residuals after sky subtraction) with a lower redshift precision
($\sim$1000 km\,s$^{-1}$).

\begin{figure}
\begin{center}
\includegraphics[width=8cm]{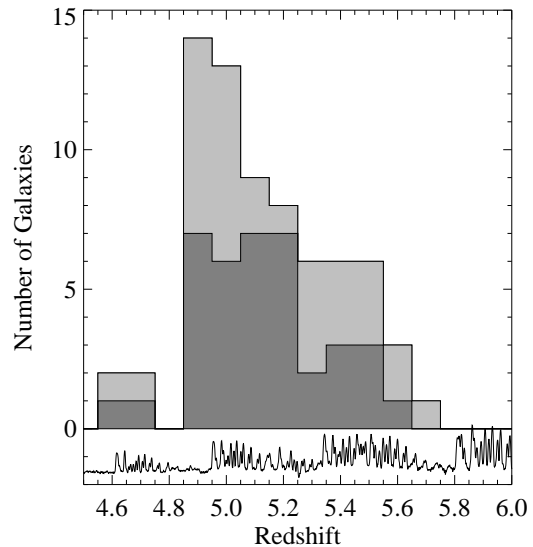}
\caption{Redshift distributions of Ly$\alpha$ line emitting LBGs (dark grey) and continuum only LBGs (plotted above the line emitters in light grey). Both grade A and B LBGs are included. Also included, below the distributions, is the sky emission spectrum at the redshift of Ly$\alpha$.}
\label{fig:redistlines}
\end{center}
\end{figure}

Figure \ref{fig:redistlines} shows the redshift distribution of the
sample divided into Ly$\alpha$ line emitting galaxies and
continuum-only Lyman break galaxies. Although more line emitters were
confirmed, the two distributions span the same redshift range. The
presence of line emission does not significantly bias the sample
completeness as a function of redshift. Of the 54 grade A spectra, 36 (67\%)
had Ly$\alpha$ line emission. The high fraction of confirmed line
emitters compared to break-only galaxies is most likely caused by the
limited signal-to-noise of the spectroscopy. The redshift of
comparatively strong line emitters is much easier to determine than
the location of a continuum break. When the whole sample of grade A and
B spectra of 70 galaxies
is considered 38 (54\%) confirmed high
redshift galaxies have Ly$\alpha$ line emission, resulting in a more
balanced detection rate for both populations. We stress that a grade B 
rating does not indicate uncertainty in the high redshift identification,
only in the precision of the derived redshift.

\section{Rest-frame UV continuum slope}
\label{sec:colours}

In the local universe, the spectral slope in the UV, $\beta$
(where $F_{\lambda}\propto\lambda^\beta$), is sensitive to the
amount of reddening, the fraction of the non-ionising continuum that
is reprocessed into infrared emission, and the metallicity of the
galaxy \citep{heckman98}. There are good correlations, with similar
significance, between the spectral slope and log L$_{IR}$/L$_{UV}$, the
strength of the relatively low ionisation metal lines from the ISM, and
the average metallicity. In addition $\beta$ is relatively insensitive
to the age of the galaxy within the first 100 Myrs, lying between $-2.6$
and $-2.0$ \citep{leitherer95}.  These correlations reflect the complex
relationships between the population of stars that emit significantly in
the non-ionising UV \citep[dominated typically by B-stars but is a strong
function of age for the youngest populations, {\it e.g.},][]{meurer97}, the
overall extinction and reddening, and the metallicity of the gas (which
influences the gas to dust ratio, the gas temperature and ionisation
state, and the metal line absorption strength). These relationships can be
used to investigate the likely properties of our distant galaxy sample.
We can also test the appropriateness of these local relationships by
making a comparison with z$\sim$3 LBGs.

\subsection{UV slope from spectroscopy}
\label{sec:imz}

Two simple average spectra were created, one for the Ly$\alpha$
emitting sources and one for the break-only LBGs, by combining the
sources with the best signal-to-noise continuum detections just
long-ward of 1216\AA~ in the rest-frame using redshifts determined by
the location of the continuum break and line emission. A
one-dimensional spectrum was extracted from each two-dimensional flux
calibrated spectrum, converted to a rest-frame wavelength scale and
combined with similar spectra of other galaxies with or without line
emission. Spectra of ten sources were combined to form the average
line-emitter spectrum and thirteen for the average break-only
spectrum. The difference in the number of spectra used for each
average affects the signal-to-noise, but not the normalisation or shape
of the final spectra. The two sets of sources had comparable redshift
distributions, average $I-$band magnitudes and signal-to-noise at
$\sim 1250$\AA~ in their rest-frame spectra. All spectra were drawn
from the same sets of masks and identically reduced with the same
calibrations; there was no way of determining {\it a priori} whether
an object would be a line-emitter or a break. While the line-emitters
were identified from the presence of both a line and break, the
identification of the break-only sources required a clear continuum
break across 1216\AA, rather than a gradual fading towards the blue and
so could potentially bias the highest signal-to-noise break-only
sample used here to those galaxies with the bluest breaks.

\begin{figure}
\begin{center}
\includegraphics[width=8cm]{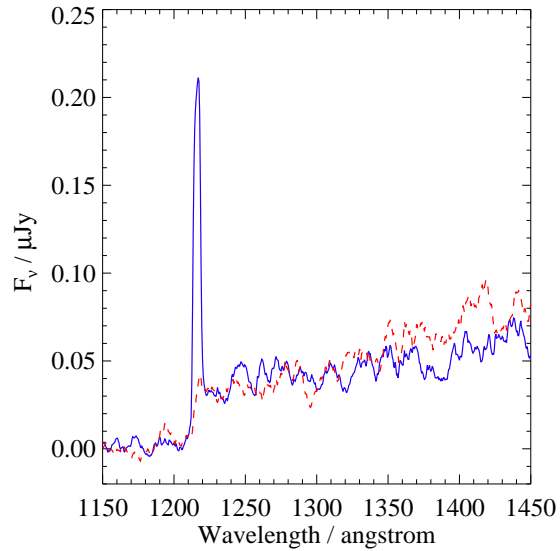}
\caption{Average spectra of line emitting (solid line) and break-only
  galaxies (dashed line) smoothed using a boxcar algorithm with a
  width of 5\AA\ The averages were made from spectra of the 10 grade A
  line emitters and 13 grade A break galaxies with the highest S/N
  continuum detections just longward of rest-frame 1216 \AA, the two
  sets of galaxies with similar redshift distributions and average
  $I-$band magnitudes.}
\label{fig:stack_spec}
\end{center}
\end{figure}

A comparison of the rest-frame UV slope for the two LBG populations is
shown in figure \ref{fig:stack_spec}. Although this comparison
indicates that break-only galaxies have a slightly redder continuum
than the line emitters (and show a difference in the opposite sense to
the potential bias noted above), this could be due to the Ly$\alpha$
absorption or broad dust features rather than the intrinsic slope of
the continuum. Without near-infrared spectroscopy, sampling longer
wavelengths, reliable measurements of the UV slope cannot be made with
this data. However, at the resolution and wavelength coverage of the
$z\sim5$ spectra presented here, both stacks are consistent with
similar stacks of line-emitting and break-only $z\sim3$ LBGs in
\citet{shapley03}.

\subsection{UV slope from photometry}
\label{sec:imp}

Previous studies have used $I-z$ colours
\citep[\textit{e.g.}][]{pentericci07}, in the absence of suitable
spectroscopy, to measure the slope of the rest-frame UV continuum of
LBG galaxies at $z=3-5$. At $z\sim5$, the ERGS $z$-band imaging (and
spectroscopy) does not provide a wavelength baseline (rest-frame
1200\AA\ to 1450\AA) sufficient for estimating the continuum slope
reliably. Instead, the longer wavelength $K_{s}$-band data is used in
combination with the $I-$band, resulting in a rest-frame wavelength
range of $\sim 1300$\AA\ to $\sim 3500$\AA. Although the near-infrared
data in this study is not deep enough to detect individual high
redshift galaxies, using the technique of stacking many objects
together enables the data to reach fainter magnitudes.

\begin{figure}
\begin{center}
\includegraphics[width=8cm]{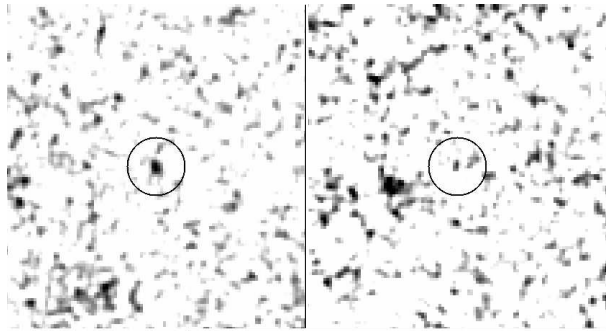}
\caption{Imaging stacks of the $K$-band images of break-only galaxies (left), Ly$\alpha$ emitters (right). Both images have the same smoothing and scaling applied.}
\label{fig:stack_lb}
\end{center}
\end{figure}

Sections of the $K_{s}$-band images, centred on the objects of interest,
were stacked together using {\it imcombine} within the {\it
IRAF} package. The mean stack was created with percentile clipping
and was then used to measure the average $K_{s}$-band flux. Two examples
of the LBG stacks are shown in figure \ref{fig:stack_lb}, the Ly$\alpha$
line emitting and break-only samples. At $z\sim5$ the Lyman break falls
into the $I$-band which affects the measurement of the continuum level. To
correct for this effect and obtain a colour which represents only the
continuum, the $I$-band magnitude of each galaxy contributing to the
stacks was corrected for the redshift dependent continuum break effect
and Ly$\alpha$ line emission.  This gives an average $I_{AB}-K_{AB}$
colour for the whole sample which can be transformed into a measure of
the UV to blue optical spectral slope.  Table \ref{tab:imk} summarizes the
stacking of the $K_{s}$-band imaging data of spectroscopically confirmed
$z\sim5$ galaxies across the ten survey fields.  We found that the
break-only galaxies are significantly redder than the line emitters,
$I_{AB}-K_{AB}=0.8$ compared with $I_{AB}-K_{AB}<0.0$, the 2$\sigma$
colour limit.

\begin{table}
\caption{Stacked $I_{AB}-K_{AB}$ colour of confirmed high redshift
galaxies, break-only galaxies and Ly$\alpha$ line emitting galaxies. The
individual $I$-band magnitudes of the galaxies have been corrected for
line emission and the redshift dependent location of the Lyman break
within the $I$-band filter. The rest-frame UV slope is described by
$\beta$ assuming $F_{\lambda}\propto\lambda^\beta$. Where a limit is
shown, the 2$\sigma$ depth of the stacked $K_{s}$-band was used.}
\begin{center}
\begin{tabular}{|c|c|c|c|c|}
\hline
Sample &  No. gals &   $I_{mean}$ &  $I_{AB}-K_{AB}$  &   $\beta$ \\ \hline
All breaks & 20 & 25.56 & $0.82\pm{0.23}$ & -1.3 \\
All lines & 22 & 25.62 & $<$0.02 & $<$-2.0 \\ \hline
All $z\sim5$ galaxies & 42 & 25.59 & $0.27\pm{0.25}$ & -1.8 \\ \hline
\end{tabular}
\end{center}
\label{tab:imk}
\end{table}

Given the degeneracy between stellar age, reddening and metallicity it
is difficult to determine the primary source of the observed colour
difference between the two subsamples. Nevertheless, we compare the
relative effects of stellar age and reddening to the $I_{AB}-K_{AB}$ colour. 

Using the \citet{calzetti00} dust models, the colour difference
between the two subsamples of at least $I_{AB}-K_{AB}\sim 0.8$ can be
produced by a small difference in the average redenning between them,
with the break-only galaxies being reddened by an extra E(B-V)$\sim
0.2$ relative to the Ly$\alpha$ emitters.  The average colours of both
samples indicate a low level of dust content, with a typical E(B-V) of
more than a few tenths, otherwise the intrinsic colours would be too
blue for even very young models with a standard IMF. The levels of
reddening are consistent with those found in photometric studies of
$z\sim 5$ LBGs \citep[\textit{e.g.}][]{verma07} and in spectroscopic
studies of $z\sim 3$ LBGs \citep[\textit{e.g.}][ finds E(B-V)=0.15 and 0.1
  for break-only and line emitting galaxies at $z\sim3$]{shapley03}.

The colour difference could also indicate an average stellar
population age difference between the two subsamples, though any age
difference is constrained by the age of the Universe at $z\sim 5$
(about 1.1 Gyr) and, more stringently, with the range of ages found in
previous photometric studies of individual $z\sim 5$ galaxies. For
example, the study by \citet{verma07} used detailed photometry out to
8$\mu$m to determine that the majority ($\sim 70\%$) of such sources
have young ($<100$ Myr) dominant stellar populations, with only a
minority older than this, typically up to 2-300 Myr.

Using composite stellar population models from \citet{maraston05} that
bracket the plausible range of star formation and metallicity for the
galaxies, we explored how age differences could affect the observed
$I_{AB}-K_{AB}$ colours. For an exponentially declining star formation rate,
with an e-folding time of 30 Myr and a metallicity of 0.05$Z_\odot$, a
colour difference of $I_{AB}-K_{AB}\sim 1$ could be obtained if the youngest
subsample was no older than a few 10s of Myr and the oldest $\sim
100$Myr older. If the youngest population was itself on average $>
100$ Myr old, the age difference increases to 200 Myr or more. For a
model where the e-folding time was 100 Myr and the metallicity 0.5
$Z_\odot$, the older subsample had to be typically 200 Myr older than
the younger for almost any plausible age of the younger population out
to 500Myr. For an unreddened $I_{AB}-K_{AB}\sim 0.8$, a source would need to be
200 or 400Myr old for an e-folding time of 30 and 100Myr
respectively. An unreddened colour limit of $I_{AB}-K_{AB}<0$ results in ages of
$<100$ and $<200$ Myr for the same models.  Associating the line
emitters with the youngest populations and the break-only galaxies
with the older, the ages of the older populations are difficult to
reconcile with the results from \citet{verma07}, given that half of
the spectroscopically-confirmed sample are break-only systems. Clearly
it is possible that a combination of differences in both stellar
population age and reddening could explain the observed colour
difference between the subsamples, but assuming the age constraints
indicated by \citet{verma07} and the colours of the \citet{maraston05}
models, the colour difference is more simply achieved through a
variation in redenning.

A difference in the intrinsic rest-frame UV continuum slope for LBGs has
been seen in other photometric and spectroscopic studies at $z=3-5$. At
$z=3$, UV slopes of $\beta=-0.7$ and $\beta=-1.1$, for break and line
emitting galaxies respectively, are measured from spectral stacks in
\citet{shapley03}. Similarly \citet{vanzella09} presents a $z\sim4$
spectral stack for the two galaxy populations which is consistent
with the UV slopes measured photometrically in \citet{pentericci07}
($\beta=-1.7$ for break galaxies and $\beta=-2.0$ for line emitters). The
results presented here follow similar trends with a photometrically
measured slope of -1.3 for break-only galaxies and $<$-2 for Ly$\alpha$
emitting galaxies.

Given that no rest-frame UV colour criteria were used in our observational
selection ({\it e.g.} no $I_{AB}-z_{AB}$ colour cut), the observed
targets were not biased to objects with blue continuum. However, the
majority of confirmed LBGs were observed to have relatively blue, flat
($\beta\sim-1.8$) rest-frame UV continuum. Only two galaxies were detected
individually in the near-IR imaging suggesting a redder continuum. As
the objects observed are bright in the UV, it is to be expected that
they have blue continuum dominated by young stellar populations. If
a sample containing objects with fainter UV luminosities were probed,
it would likely contain a higher proportion of older and perhaps more
reddened galaxies.

\subsection{The Metallicity of z$\sim$5 LBGs}
\label{sec:metallicity}

\begin{figure}
\begin{center}
\includegraphics[width=8cm]{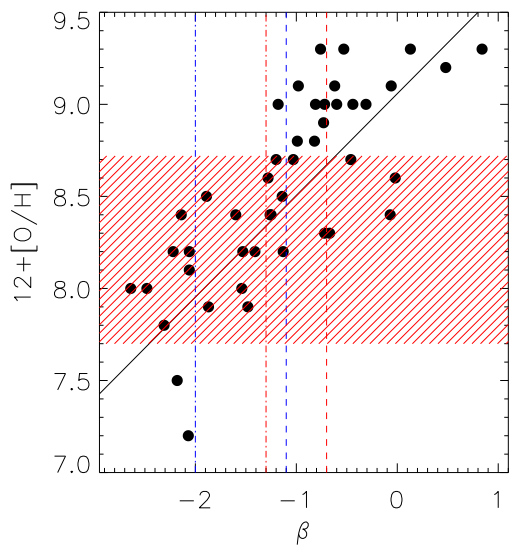}
\caption{Plot of the UV spectral slope, $\beta$, versus the metallicity.
The black dots show the spectral slope and metallicity of local galaxies
from \citet{heckman98}, the dot-dashed vertical lines indicate the
spectral slope of the galaxies with breaks only in our sample (red)
and for the whole sample with spectroscopic redshifts (blue), and the
dashed vertical lines indicate the spectral slope of the galaxies with
Ly$\alpha$ mostly in absorption (red) and galaxies with Ly$\alpha$
predominately in emission (blue) from \citet{shapley03}. The red horizontal
region shows the area encompassing the metallicity estimates from the
literature (see text for details).}
\label{fig:Zbeta}
\end{center}
\end{figure}

\citet{heckman98} found correlations between the spectral slope and log
L$_{IR}$/L$_{UV}$, the strength of the relatively low ionisation metal
lines from the ISM, and the average metallicity for nearby galaxies. In
Figure~\ref{fig:Zbeta}, we plot the spectral slope, $\beta$, versus the
metallicity, 12+log[O/H], from the \citet{heckman98} sample of local
galaxies and their best fit to these data. Overplotted are the estimates
of the spectral slope of $z\sim3$ LBGs from \citet{shapley03} and the
range of metal abundances determined for $z\sim3$ LBGs \citep{pettini01,
nesvadba06, nesvadba08, mannucci09}. Two values of the spectral slope
from \citet{shapley03} were used, a steeper slope for the Ly$\alpha$
emission line galaxies and a redder slope for the galaxies where
Ly$\alpha$ is in absorption.

We see that, in general, $\beta$ for the $z\sim3$ LBGs predicts a higher
metallicity than has been observed.  The number of good metallicity
estimates is low and the comparison may suffer from biases (such as
slope estimates weighted towards the brightest galaxies, which have
populations that are more evolved, or the emission line estimates are
weighted towards galaxies of lower metallicities and dust content on
average due to the influence of metallicity on the gas temperature,
effecting the line strengths of rest frame optical lines).
\citet{heckman98} found that galaxies brighter in the optical and
with higher (approximate) bolometric luminosities have bluer spectral
slopes.  Realising that in this work  we are discussing the brightest LBGs at z$\sim$5, the spectral slopes suggest that the $z\sim5$ LBGs
are roughly a factor of three lower in metallicity than the $z\sim3$
LBGs and are approximately 1/5-1/10th of solar metallicity but this
estimate may be biased towards galaxies of higher metallicities.

About 10-20\% of the LBGs at $z\sim3$ have stellar ages consistent
with forming at $z\sim5$.  The growth in the number density and in the
typical mass of a LBG over this period are also roughly a factor of
10.  This may suggest a direct evolutionary connection between the two
populations \citep[cf. paper I,][]{verma07, shapley03,
  steidel99}. Moreover, in an analysis of $z\sim3$ LBGs,
\citet{mannucci09} find that galaxies with masses similar to those
expected for this sample \citep{verma07} have metallicities consistent
with our estimates using the measured $z\sim3$ $\beta$ and the local
relationship between $\beta$ and metallicity. For such systems,
\citet{mannucci09} have estimated gas fractions of 80\% or more. For
the oldest systems at $z\sim3$, which are also the most massive
\citep[see also][]{erb06}, the gas fractions are estimated to be much
less ($\sim$20\%).  Therefore, although beyond the scope of the
present paper, it is possible to relate the more massive, older, more
metal-rich populations of LBGs at $z\sim3$ to the population we
observe at $z\sim5$.

\section{Lyman $\alpha$ Line Properties}
\label{sec:ew}

The Lyman $\alpha$ line is the only emission or absorption line
straightforwardly detectable in the observed optical in a reasonable
time frame (four hours using a 8m telescope) in the spectra of LBGs at
$z\sim5$. A small number of observations of the continuum of bright
or lensed $z\sim5$ galaxies have reached sufficient depths to probe
interstellar absorption features in the rest-frame UV continuum
\citep{ando04,ando07,dow05}.

\subsection{Lyman $\alpha$ at $z\sim5$}

The Ly$\alpha$ line flux of each galaxy was measured within a square
aperture centred on the line emission in the two dimensional spectrum. The
size of the aperture was tuned to the apparent spatial and spectral
FWHM of each galaxy. The flux was integrated over the wavelength
range within the aperture, yielding a final measurement in units of
ergs\,cm$^{-2}$\,s$^{-1}$. Line fluxes ranged from $2.6\times10^{-18}$
to $7.0\times10^{-17}$\,ergs\,cm$^{-2}$\,s$^{-1}$ (luminosities between
$10^{42}$ and $10^{43}$ erg s$^{-1}$) with a typical 3$\sigma$ line flux
limit of $2.9\times10^{-18}$\,ergs\,cm$^{-2}$\,s$^{-1}$ for an unresolved
emission line. At the low redshift end of the
sample there is a lack of strong emission lines as a result of the line
flux contributing to the $R$-band magnitude, reducing the $R-I$ colour
and causing the object to drop out of the original colour selection. From
the $z>5$ Ly$\alpha$ equivalent width distribution potentially 10-30\% of
line emitters could be missed at $z<5$ given the influence of Ly$\alpha$
emission on the $R_{AB}-I_{AB}$ colour \citep{stanway08b}

\subsection{Equivalent Width of Ly$\alpha$}

\begin{figure}
\begin{center}
\includegraphics[width=8cm]{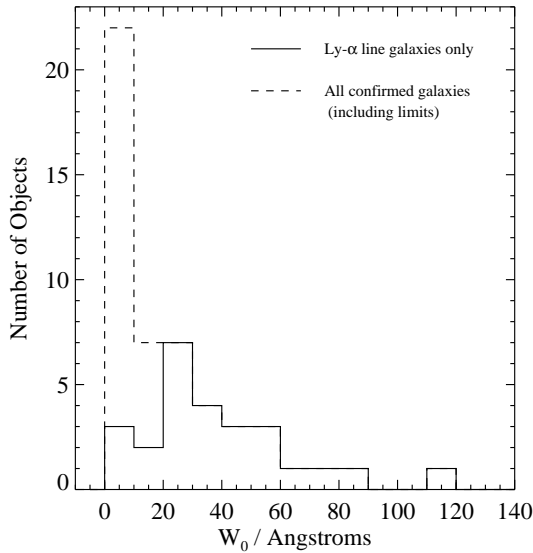}
\caption{Distribution of the rest-frame Ly$\alpha$ equivalent width
for line emitters (solid line) and when 3\,$\sigma$ upper limits from
spectroscopically-confirmed continuum break galaxies are included as
described in the text (dashed line). Continuum flux levels were
measured from broadband photometry.}
\label{fig:restew}
\end{center}
\end{figure}

As the observed galaxies are faint ($26.3>I_{AB}>25.0$), the continuum
emission detected spectroscopically has low signal-to-noise. As such,
equivalent width measurements (effectively the ratio of the Ly$\alpha$
line flux to continuum flux density) given here use the broadband
photometry of individual galaxies to estimate their continuum
flux. Measured line flux was removed from the 2$''$ aperture $I$-band
magnitude to determine each galaxy's continuum flux level, accounting
for a redshift dependent correction due to IGM absorption shortward of
Ly$\alpha$. A limit on the Ly$\alpha$ equivalent width of the
break-only galaxies was measured by introducing an artificial line
feature with a flux equivalent to the 3\,$\sigma$ variations in the
local background at the break wavelength. This simulated emission line
was placed at the continuum break of the spectrum and used to estimate
an upper limit on the observed equivalent width limit which accounted
for the noise properties of each spectrum. The derived rest-frame
equivalent widths are shown in figure \ref{fig:restew}. The strongest
line detected has a rest-frame equivalent width $W_0=112\pm18$\,\AA,
the weakest detected line had $W_0=6\pm2$\,\AA.

While figure \ref{fig:restew} presents the distribution of Ly$\alpha$
line strengths for sources with spectroscopically confirmed redshifts,
this is an incomplete picture. To develop a more complete
understanding of the equivalent width distribution at $z\sim5$, it is
necessary to consider those sources which were spectroscopically
observed, but did not yield a reliable redshift determination or
recognisable spectral features. Unsurprisingly, these sources
outnumber those with spectroscopic confirmations and must comprise a
mixture of faint galaxies, sources at intermediate brightness with
weak or absent emission lines, and interlopers contaminating the
photometric sample.

In Paper I we determined that $\sim$80\% of our refined photometric
sample is likely to lie at $z>5$. Considering a similar subset of
sources with reliable multiband photometry, selected as strong high
redshift candidates and placed on the spectroscopic masks, we identify
57 sources for which no redshift was determined, of which $\sim45$ are
expected to be genuine high redshift galaxies, but whose emission
lines may lie anywhere in the observed spectrum. In the following we
estimate the contribution of the upper limits to any Ly$\alpha$
emission from these sources to the complete equivalent width
distribution.

\begin{figure}
\begin{center}
\includegraphics[width=8cm]{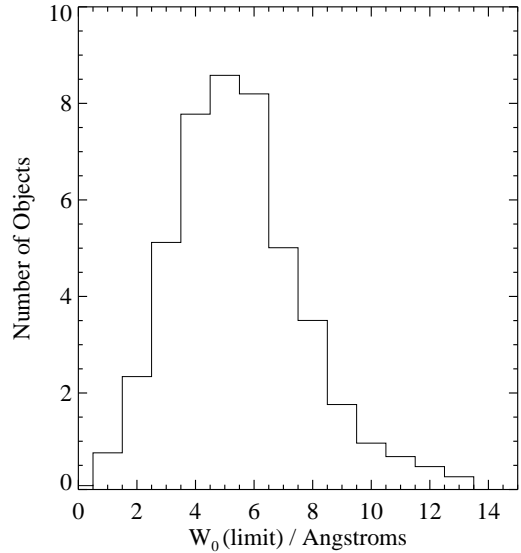}
\caption{Estimated distribution of rest-frame Ly$\alpha$ equivalent
width upper limits for observed sources without distinguishable
features in their spectra. This distribution was determined by
convolving 3\,$\sigma$ upper limits measured from the spectra as a
function of redshift with the redshift distribution calculated for the
sample, as described in the text.}
\label{fig:restew2}
\end{center}
\end{figure}

We calculate a probability distribution of upper limits for this
sample as follows.  For each source in the sample, we determine a
3\,$\sigma$ upper limit on the flux of an undetected emission line as
a function of redshift, accounting for wavelength-dependent noise in
the spectrum. This is converted to a limit on the rest-frame
Ly$\alpha$ equivalent width as a function of redshift, using the
$I$-band magnitude of the source to determine a continuum level (and
accounting for IGM absorption shortwards of the proposed emission
line, again as a function of redshift). We weight the probability of
the line occurring at each redshift by the calculated redshift
distribution function of this sample (figure \ref{fig:redist}),
yielding a probability distribution for the 3\,$\sigma$ upper limit on
the equivalent width for each source. The distribution is then scaled
by 80\% to produce an estimated contribution to the total sample
equivalent width distribution. The resultant distribution of line
upper limits is shown in figure \ref{fig:restew2}.  The typical
3\,$\sigma$ upper limit on a line detection is $5\pm2$\,\AA, with the
small amplitude of the tail to higher limits suggesting that our
sample is highly complete for line emitters down to our flux limit.

Any interpretation of the Ly$\alpha$ line is challenging given its
complex and ambiguous nature.  The observed line profile is affected
by resonant scattering within the galaxy by interstellar HI. This
introduces influences from the geometry, kinematics and dust content
of large-scale outflows.

How does the $z\sim 5$ equivalent width distribution compare to that at lower redshifts? At $z\sim3$, \citet{shapley03} determined the equivalent width
distribution of Ly$\alpha$ for nearly 1000 galaxies. They find that
25\% of $z\sim3$ Lyman break galaxies have a Ly$\alpha$ equivalent
width greater than 20\AA. Of galaxies in our
spectroscopically-confirmed sample at $z\sim5$, 55\% have
$W_0>20$\AA. However, when we consider the larger sample of
spectroscopically-observed high redshift candidate sources and the
estimated contribution of unconfirmed galaxies in figure
\ref{fig:restew2} is taken into account, this fraction falls to
$22\pm4$\%. 

Previous work at lower redshift including \citet{shapley03} similarly
did not include a correction for unconfirmed sources.  While the
confirmed fraction at $z\sim3$ (63\%) in that work is significantly
larger than that in this sample, and included sources with Ly$\alpha$
seen in absorption, it is likely that an essentially unknown fraction
of the unconfirmed sources would contribute to the Ly$\alpha$
equivalent width distribution in the same manner as those discussed
above. Consequently, the quoted fraction of sources with high
equivalent widths at $z\sim 3$ is an upper limit, but nevertheless,
even if all unconfirmed sources in the \citet{shapley03} sample are at
$z\sim 3$, the high equivalent width fraction drops to $\sim 18\%$,
consistent with our $z\sim 5$ statistics.

Interestingly, this survey is not alone in finding a large fraction of
{\it confirmed} high redshift galaxies with a large Ly$\alpha$ equivalent
width. A long tail of sources with high equivalent widths is seen by
\citet{dawson07} at $z=4.5$, \citet{hu04} at $z=5.7$ and
\citet{stanway07} at $z\sim6$, although these studies discuss only
sources with spectroscopic confirmation.  In contrast \citet{dow07} do
not observe such a tail based on 6 line emitting galaxies at
$z\sim6$. At present, all of these studies suffer from low
spectroscopic completeness and small number statistics. The current
study is the first at high ($z>5$) redshift with sufficient number
counts and understanding of selection functions to make a meaningful
comparison of the equivalent width distribution to that seen at lower
redshifts.

\begin{figure}
\begin{center}
\includegraphics[width=8cm]{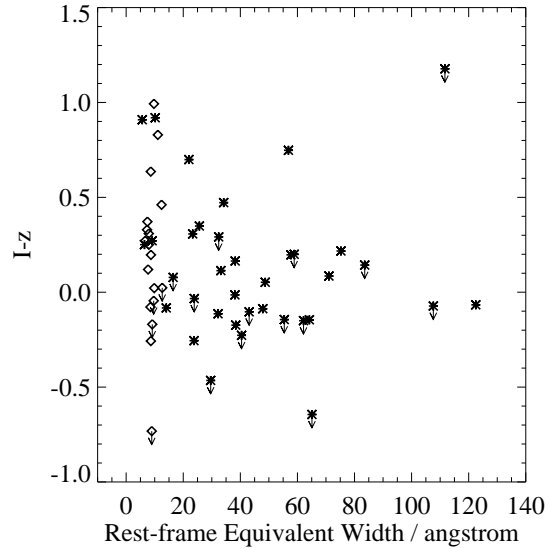}
\caption{Rest-frame equivalent width of Ly$\alpha$ as a function of
$I_{AB}-z_{AB}$ colour. Diamonds represent equivalent width limits of
the break-only galaxies as described in the text. The colour has been
corrected for line emission and IGM absorption effects. Arrows indicate
a non-detection in the $z$-band and hence a colour limit.}
\label{fig:ewcont}
\end{center}
\end{figure}

The largest equivalent widths seen at $z\sim3$ are almost double
those at $z\sim5$. Given their rare occurrence, their absence in the
higher redshift sample may simply be an effect of small number
statistics. \citet{shapley03} also found that the UV continuum slope
became bluer as the equivalent width of Ly$\alpha$ increased in
stacked samples of LBGs at $z\sim3$. Figure \ref{fig:ewcont} shows the
distribution of $I_{AB}-z_{AB}$ colour, a rough indicator of the UV
continuum slope once corrected for line emission, with the rest-frame
equivalent width of Ly$\alpha$. Although a comparably clear trend is
not seen, sources with equivalent widths of more than 30\AA\ appear to
have bluer colours than those with lower equivalent widths. There are
more $z$-band detections resulting in redder colours for sources with
W$_o<$30\AA. However, given the small number of sources this is not
highly statistically significant.

The resolution of the spectroscopy was such that only the strongest Ly$\alpha$ lines were barely resolved, so little can be said about the distribution of line widths and line shapes in the sample; higher resolution spectroscopy is required.

\section{Morphology of $z\sim5$ LBGs}
\label{sec:specmorph}

\begin{figure}
\begin{center}
\includegraphics[width=8cm]{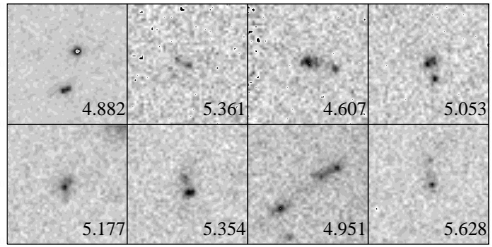}
\caption{Eight examples of spectroscopically confirmed galaxies with
multiple components. Each box is $5''\times5''$ and the images were
taken using the HST/ACS camera and the F814W filter.Redshifts are indicated in the bottom right-hand corners}
\label{fig:specinter}
\end{center}
\end{figure}

HST/ACS imaging using the F814W filter \citep[$I$-band,][]{desai07} were
used to investigate the morphology of the confirmed high redshift LBGs
\citep[see][for details]{douglas09}.  Of the spectroscopically confirmed
LBGs, 20 individual sources have deep HST imaging. Within this sample,
14 LBGs have multiple components, four show irregular morphology, one is
compact and unresolved and the final object is compact but surrounded by
diffuse emission. Examples of LBGs with multiple components or irregular
morphology are shown in figure \ref{fig:specinter}. The morphology of
individual spectroscopically confirmed high redshift galaxies is similar
to that of the larger photometrically selected sample of $z\sim5$ Lyman
break galaxies \citep{douglas09}. A similar mean half-light radius of
0.15$''$ (or 0.9kpc) and distribution of sizes was observed for the
two samples, and are in agreement with other high resolution studies
\citep[\textit{e.g.}][]{bremer04,bouwens06}. The number of components identified
in the multiple systems ranged from two to four, however the systems
may contain additional fainter components that are not detectable above
the surface brightness limits of the images. The mean distance between
components is 0.84$''$ (5.3\,kpc) with distances observed between 0.2$''$
and 2.5$''$ (1\,kpc and 16\,kpc) compared to a seeing disk of typically
0.7$''$ in the groundbased imaging.

Two systems are resolved in the ground-based imaging with component
separations of 0.8$''$ and 1.0$''$, however further components have
been identified in the deeper, high resolution HST images. Within each
system, the previously identified components have similar optical colours
with no detections in the nIR imaging. Both systems are marginally
resolved in the spectroscopy showing two sources. One system has two
faint Ly$\alpha$ emission lines and the other has a line emitter and a
break-only galaxy. The component redshifts measured from these features
are within 1000\,km\,s$^{-1}$, which, although high, does not rule out a
connected system as redshifts measured using Ly$\alpha$ emission or the
Lyman break are not necessarily equal to the systematic redshift of the
galaxy. Velocities of up to 1000\,km\,s$^{-1}$ between Ly$\alpha$ emission
and interstellar absorption lines have been observed in $z\sim3$ LBGs.

A higher incidence of complex morphology in the
spectroscopically-confirmed sample is observed compared to the larger
photometric sample where $\sim$30\% of galaxy candidates had more than
one component or irregular morphology. This effect has not been
observed in other spectroscopically confirmed samples
\citep{vanzella05,vanzella09}. However, given the small number of
spectroscopically confirmed z$\approx$5 LBGs and differences in
selection and image depths, these numbers are statistically consistent
with other studies \citep{vanzella05,vanzella09,conselice09}.

\subsection{The nature of the multiple components}

The relatively high fraction of distant LBGs that have multiple
components may tell us important information about their nature of their
star-formation and its triggering mechanism.  The high fraction of pairs
or triplets suggests that the mergers and instabilities that form massive
regions of intense star-formation within larger underlying structures
may play an important role in $z\sim5$ LBGs. The observed faint structures
surrounding these sources might be indicative of mergers, but could also
be structure within the light distribution of a single larger object. There is
currently no direct evidence to favour either scenario.

An appropriate way to understand the nature of z$\sim$5 LBGs may be
through analogy with a local population. \citet{overzier08} have
studied nearby analogues of LBGs -- Lyman Break galaxy analogues or
LBAs. These LBG analogues were chosen to have properties similar to
$z\sim3$ LBGs, namely, L$_{FUV}>10^{10.3}$ L$_{\sun}$ and
I$_{FUV}>10^{9}$ L$_{\sun}$ kpc$^{-2}$. Thus they mimic the high UV
luminosities and intensities observed in $z\sim3$ LBGs. Overall, the
LBAs appear to be similar to the $z\sim5$ LBGs, perhaps more so than
the $z\sim3$ LBGs they were choosen to mimic.  Their total stellar
masses are between 10$^9$ and 10$^{10}$ M$_{\sun}$, close to the
typical masses of bright (brighter than L$^{\star}$) $z\sim5$ LBGs
\citep{verma07}, and low metallicities and similar spectral slopes (
\S~\ref{sec:imz}, and \S~\ref{sec:imp}, \S~\ref{sec:metallicity}).
Their UV luminosities are typically a factor of a few to 10 times lower
($\lambda$P$_{\lambda}$ for $\lambda$=1900\AA) than our sample. Both
the small scale clustering properties and the internal kinematics of
the LBAs suggest a merger or interaction origin for the intense
star-formation observed in these systems \citep{b-z09a, b-z09b}.
Given the apparent similarities, a detailed comparison seems
warranted.

\citet{overzier08} find that the clumps of star-formation, or SSBs, in
the LBAs, have mass surface densities of 10$^{2-3}$ M$_{\sun}$
pc$^{-2}$ over radii of $\sim$100 pc, with typically one to more than
10 large SSBs per galaxy. Such systems would be unresolved at z$\sim$5
in any of our datasets. The half-light radii of the galaxies
themselves in the UV or U-band are typically around one kpc. A
comparison of this mass surface density with the average surface
densities of the \citet{verma07} sample (comparable to LBGs studied
here) gives a similar average mass surface density
($\approx$6$\times$10$^{2}$ M$_{\sun}$ pc$^{-2}$ over radii of $\sim$1
kpc or the half light radii of $z\sim5$ LBGs). In the LBAs, about
30-80\% of the total inferred star-formation occurs in such
super-starburst (SSB) regions.  Thus, the $z\sim5$ LBGs could be
thought of as a ``galaxy wide'' super-starburst region, with a high
density of star-formation that must contribute nearly 100\% of the
inferred star-formation rates (which would also explain the higher UV
luminosities for the $z\sim5$ LBGs compared to their local analogues).
In addition, \citet{overzier08} infer ages of the SSBs of about 10-100
Myrs, but formally regard these as upper limits due to complex
extinction in the UV and due to the (unknown) contribution of older
stars.  Interestingly, \citet{verma07} derived similar ages from model
fits of the bandband photometry of $z\sim5$ LBGs.  This is all
consistent with the picture that the $z\sim5$ LBGs are possibly a
collection of galaxy wide SSBs with a contribution to the total
star-formation of nearly 100\%.

If the LBGs studied here are analogous to the LBAs, with spatial
distributions over one to 5\,kpc, with typical separation between components 
of around 1\,kpc, we would only expect to distinguish some of the
separate SSB clumps in HST images \citep[{\it e.g.}, Fig.~7
  in][]{overzier08}.  For the spectroscopically-confirmed galaxies
with deep HST imaging, we find 12 that are clearly multiple with an average
projected separation of 5.3\,kpc and range from separations of one to
16\,kpc. For the total photometric sample with deep HST imaging, 14
out of 95 are multiple systems and 16 out of 95 show disturbed
morphology \citep[modulo the contamination which should be
  low][]{douglas09}.  Taken at face value, the high fraction of
multiple or disturbed systems suggests a bias towards
spectroscopically confirming such objects. This suggests that the
galaxies with multiple components of star formation allow for channels
to be cleared of neutral hydrogen, which then provide a long mean free
path for the Ly$\alpha$ photons to escape.  Since these galaxies
likely drive outflows \citep{verma07}, the nature of the ISM may be
such to allow the efficient escape of Ly$\alpha$ photons in analogy
with local starburst galaxies \citep{marlowe99, meurer92, atek08,
  hayes07}.

\section{$z\sim5$ Surface Density}
\label{sec:surface}

The average observed surface density of the most reliable photometric
sample of galaxies was 0.93 per arcmin$^2$ down to $I_{AB}=26.3$ (see
table 2 in paper I), comparable to the surface density observed by
\citet{bouwens07} to the same flux limit from space-based imaging
surveys. However, there is a large variation in the surface density of
LBG photometric candidates and spectroscopically confirmed objects
between individual fields. Table \ref{tab:all} show this variation in
the number of spectroscopically confirmed objects. As expected, the
number of confirmed galaxies in each field broadly correlates with the
number of photometric candidates. A factor of 8 is observed between
the richest and poorest LBG candidate fields and those with the most
and least spectroscopic identifications, beyond that expected from
cosmic variance predictions \citep{trenti08}.

\begin{table}
\caption{All confirmed high redshift Lyman break galaxies per field according to the quality of the spectra and the identifying feature.}
\begin{center}
\begin{tabular}{|c|c|c|c|c|c|}
\hline
\bf Field &  \multicolumn{2}{|c|}{\bf A }  & \multicolumn{2}{|c|}{ \bf B} & Total \\
& Ly$\alpha$ & break & Ly$\alpha$ & break &\\ \hline
J1037.9-1243 &   & 2 &   & 1 & 3 \\ 
J1040.7-1155 & 7 & 2 & 1 & 1 & 11  \\
J1054.4-1146 &   & 1 &   & 2 & 3 \\
J1054.7-1245 & 11& 4 & 1 & 8 & 24 \\
J1103.7-1245 & 3 &   &   & 2 & 5  \\
J1122.9-1136 & 2 & 1 &   &   & 3  \\
J1138.2-1133 & 5 &   &   & 1 & 6  \\
J1216.8-1201 & 2 & 2 &   &   & 4 \\
J1227.9-1138 & 2 & 2 &   &   & 4  \\
J1354.2-1230 & 3 & 2 &   &   & 5  \\ \hline

\end{tabular}
\end{center}
\label{tab:all}
\end{table}

One possible explanation of the observed variation is the changing effective area of each field. However, with variations in area of less than 10\% and no correlation with the density of candidates and confirmed LBGs suggests that this cannot account for the differences. 

Alternatively, the presence of foreground clusters in each field (the
original subject of the majority of the imaging by the ESO large
program EDisCS, \citep{white05}) could be distorting the source plane
and affecting the probed volume \citep{clowe06}. Given the steep
nature of the faint-end of the $z\sim5$ LBG luminosity function
\citep{douglas09,bouwens07}, the number density of candidates could
increase significantly with the flux boosting effect of lensing. As
the area of each field affected by strong lensing is small, this is
unlikely to be the cause of the observed number variation. There are
also no objects observed with highly distorted or elongated
morphologies, suggesting strong lensing in the deep images. Although
the effects of weak lensing cover a larger proportion of the field, the
subsequent distortion effects are small and do not significantly alter
the probed volume. There is also no correlation seen between the
number of candidates and confirmed LBGs and the velocity dispersion,
and hence mass, of the identified foreground clusters. Therefore it is
unlikely that the lensing effects of foreground structures are causing
the observed differences in LBG surface densities. It is far more
likely that these fields are probing truly overdense areas of the
universe. The most compelling evidence for this is the variation from
a smooth redshift distribution seen in the spectroscopic results of
two of the individual fields and is discussed in the next section.

\subsection{Spectroscopically confirmed overdense fields}

Out of the ten survey fields, two showed strong spectroscopic evidence of large scale structure. In fields J1040.7-1155 and J1054.7-1254 there are clear spikes in the redshift distributions at $z=5.15$ and $z=5.0$ respectively, with six or more galaxies lying within a 0.1 redshift bin, which is equivalent to a co-moving distance of around 50Mpc. The redshift distributions for these two fields are shown figures \ref{fig:red1040} and \ref{fig:red1054} for J1040.7-1155 and J1054-7.1245 respectively. The red dashed line indicates the distribution of grade A galaxies, the blue dotted line shows the distribution of all grade A and B galaxies, and the black line shows the night sky emission spectrum for the redshift of Ly$\alpha$. The total redshift distribution of all confirmed galaxies predicts, at most, 1.3 galaxies per 0.1 redshift bin for an individual field.

\subsubsection{J1040.7-1155}
\label{struct1040}

\begin{figure}
\begin{center}
\includegraphics[width=8cm]{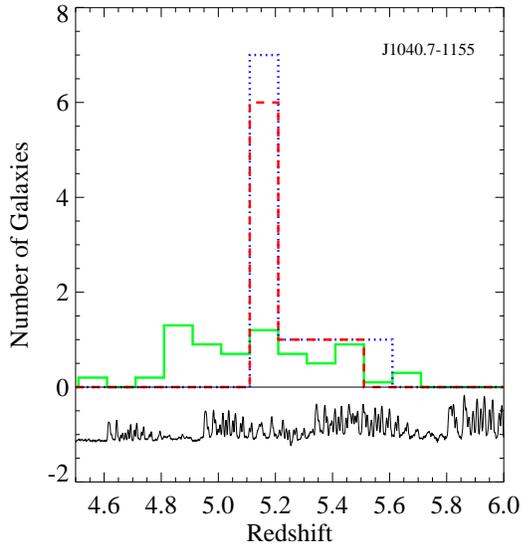}
\caption{Field J1040.7-1155: The redshift distribution of grade A confirmed galaxies is shown by the dashed red line and the distribution of all confirmed galaxies (grade A and B) is shown by the dotted blue line binned with the same parameters as the individual field data. The solid green line is the total redshift distribution averaged over the ten survey fields. The sky emission spectrum offset from zero at the redshift of Ly$\alpha$ is shown by the solid black line.}
\label{fig:red1040}
\end{center}
\end{figure}

\begin{figure}
\begin{center}
\includegraphics[width=8cm]{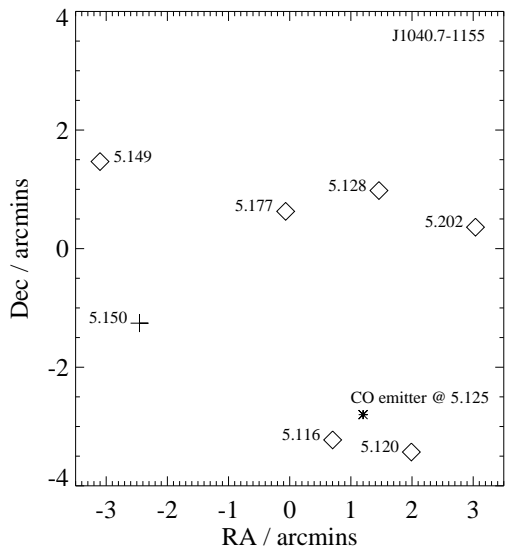}
\caption{Spatial distribution of spectroscopically confirmed grade A galaxies (diamonds) and grade B galaxies (crosses), which populate the redshift spike, across the field J1040.7-1155. Coordinates are relative to the field centre. The asterisk denotes the position of the \citet{stanway08a} CO source discussed in the text.}
\label{fig:redxy1040}
\end{center}
\end{figure}

This field contains the third largest photometric sample of high redshift galaxy candidates out of the ten survey fields, with 25 candidates in the photometric sample (paper I) and 11 high redshift spectroscopic confirmations (grade A and B). Given the size of the photometric sample, a similarly high number of spectroscopic confirmations would be expected. In the redshift distribution there is a clear peak of objects between $5.11<z<5.21$, a 6$\sigma$ overdensity when compared to the spectroscopic statistics of the survey as a whole (with the boundaries of the redshift bins tuned to give the largest peak). Figure \ref{fig:red1040} shows this peak. Only line emitters with redshift determinations accurate to within 300-400\,km\,s$^{-1}$ populate the redshift spike strengthening its identification as a real feature. It cannot be explained as an artifact created by the uncertainty in a Lyman break overlying skyline residual noise. Figure \ref{fig:redxy1040} shows the spatial distribution of the galaxies in the redshift spike across the field. Although the galaxies are spread across the field and are not concentrated in one region, it is still probable that they are part of the same structure. The closest pair of grade A galaxies in the field are 7.3\,Mpc apart, with a furthest extent of 43.7\,Mpc for the spectroscopically confirmed grade A members of the cluster, where the distances are co-moving in three dimensions. The co-moving dimensions of the surveys fields are typically 27 by 27\,Mpc in the plane of the sky and probing a redshift interval of one, equivalent to 525\,Mpc.

The spectra of the galaxies constituting this redshift spike are typical of the line emitting sample as a whole. Similarly, the morphologies and colours of the galaxies within the redshift spike are typical of the sample.

\subsubsection{J1054.7-1245}
\label{struct1054}

\begin{figure}
\begin{center}
\includegraphics[width=8cm]{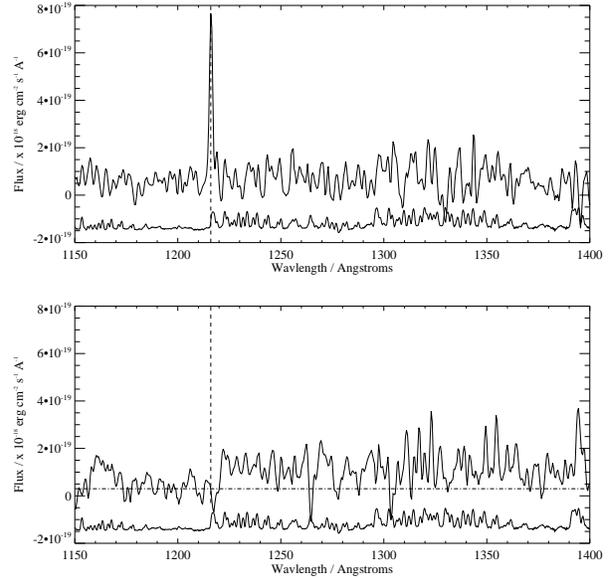}
\caption{One dimensional spectra of two LBGs separated by 1.1$''$ or 7kpc on the sky. The sky spectrum at this redshift is plotted below each spectrum and the location of Ly$\alpha$ at $z=4.951$ is shown.}
\label{fig:double}
\end{center}
\end{figure}

\begin{figure}
\begin{center}
\includegraphics[width=8cm]{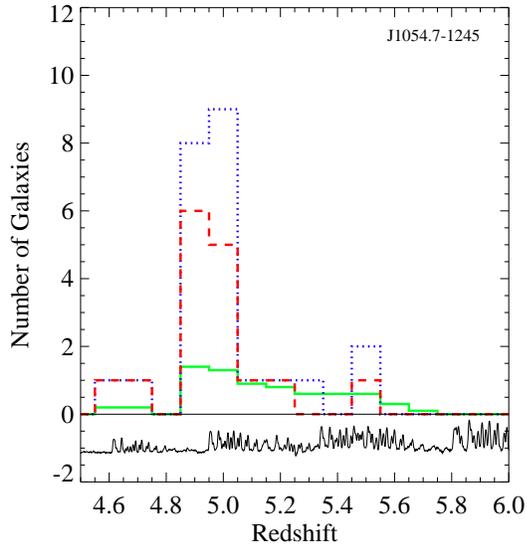}
\caption{Field J1054.7-1245: The redshift distribution of grade A confirmed galaxies is shown by the dashed red line and the distribution of all confirmed galaxies (grade A and B) is shown by the dotted blue line. The solid green line is the total redshift distribution averaged over the ten survey fields binned with the same parameters as the individual field data. The sky emission spectrum offset from zero at the redshift of Ly$\alpha$ is shown by the solid black line.}
\label{fig:red1054}
\end{center}
\end{figure}

This field contains the most photometric high redshift candidates (45
candidates in paper I), and was expected to produce the largest number
of spectroscopic confirmations. Within the spectroscopic
identification of 24 high redshift galaxies (grade A and B), there is
significant spike at $z=5.0$ within $\delta z=0.1$, containing six
times more galaxies than expected over such a redshift range (grade A
only), see figure \ref{fig:red1054}. The spike includes a close pair
with a separation large enough to be individually spatially resolved
in the spectrum. This pair was identified as a Ly$\alpha$ emitter and
break-only galaxy as shown in figure \ref{fig:double}. When both
grades of spectroscopic confirmations of high redshift galaxies are
included, this overdensity contains nine galaxies. If we include
objects between $z=4.95$ and $z=5.15$, i.e. $\delta z=0.2$, 17 grade A
and B galaxies are included. However this redshift range is much
larger than that expected for a cluster of galaxies or any other bound
structure. The broad nature of this spike may suggest we are observing
a filamentary structure end-on. Again the overdensity is mainly
determined by line emitters, a reflection of the greater fraction of
line emitting galaxies confirmed compared to break-only galaxies in
this field.

\begin{figure}
\begin{center}
\includegraphics[width=8cm]{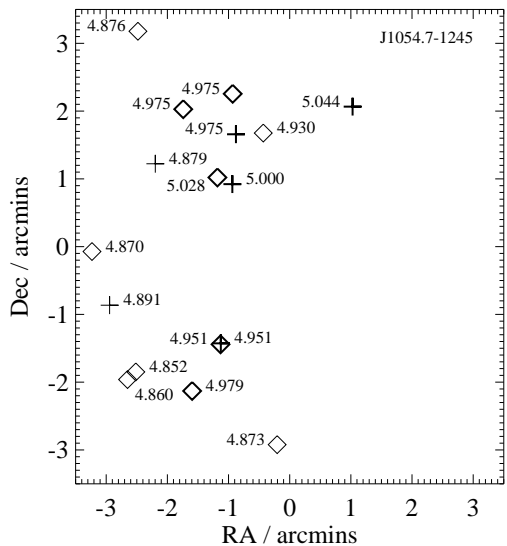}
\caption{Spatial distribution of spectroscopically confirmed grade A galaxies (diamonds) and grade B galaxies (crosses), which populate the two 0.1 redshift spikes (distinguished by bold symbols), across the field J1054-7.1245. Coordinates are relative to the field centre.}
\label{fig:redxy105412}
\end{center}
\end{figure}

The galaxies in the redshift spike are spatially distributed over a smaller fraction of the field than those in J1040.7-1155. They are clustered to one side of the $7\times7$ arcminute field, as shown in figure \ref{fig:redxy105412}. The closest grade A galaxy pair has a co-moving separation of 670\,kpc (excluding the very close pair), with the overdensity of galaxies covering 27\,Mpc (co-moving distance between furthest grade A confirmed galaxies in three dimensions). Again, the spectra, colours and morphologies of the galaxies within the redshift spike are typical of the sample as a whole.

\subsection{The Nature of the Large Scale Structure}
\label{sec:lss}

As discussed above, two of the ten survey fields show evidence of large scale structure in the form of spikes in their redshift distributions. Such overdensities could be a forming cluster, sheet or filament being probed in these fields. Given that across the survey as a whole zero to one sources per 0.1 in redshift are expected, the identification of two fields with very clear overdensities of many times this expected number is highly significant. However, given the stochastic properties of LBGs \citep{verma07}, the nature of these overdensities is not immediately clear.

As the typical star-formation lifetime of high redshift LBGs is a few tens Myrs \citep[assuming the majority of the galaxies within the overdensities are young,][]{verma07}, the apparent coordination or synchronisation of these short-lived star-bursts over the entire field needs explanation. Given the distances between the galaxies within the redshift overdensities, it is highly unlikely that these galaxies are dynamically bound or interacting sufficiently to have a common starformation trigger. Assuming the young ages of these galaxies predicted by \citet{verma07}, $<$45\,Myrs, it is likely that the galaxies have only travelled 50\,kpc, outside the Hubble flow, in their lifetime (assuming the galaxy is travelling at 1000\,kms$^{-1}$). This suggests that if two galaxies are more than 100\,kpc apart, it is unlikely that they have interacted. With the closest pair of galaxies (in field J1054.7-1245) separated by 670\,kpc (co-moving distance in three dimensions), the star formation observed cannot be connected. This limited interaction region is reflected in the high resolution imaging of the HST data. It was found that many objects detected in the ground based data were in fact multiple systems, at most 16\,kpc apart and are subsequently close enough to potentially have star formation triggered by the mutual interaction.

The apparent synchronisation of the relatively short-lived starbursts
is likely to be an illusion for the above reasons. A more plausible
explanation for the overdensity of objects undergoing a relatively
short-lived phenomenon is that there is far more baryonic material in
the large-scale structure that hosts these galaxies than is visible in
the rest-frame UV. At any one time a small fraction of this material
is visible as unobscured UV-bright starburst episodes lasting a few
tens of Myrs. If one were to wait a few tens of millions of years and
re-observe the region, the large scale structure would still be marked
out by an overdensity of UV-luminous starbursts, but probably
different ones to those seen today. Therefore, there is no
synchronisation of the starbursts, just a high probability that
several regions of the large-scale structure will be `on' at any one
time. One way to try to detect this `dark' material connecting the
observed systems, is to look for line or continuum emission, such as
carbon or CO emission lines, from gas within the structure, in the
millimetre and submillimetre regimes. Such an experiment was carried
out in a 2 arcminute diameter region centred between two LBGs with
concurrent redshifts in one of the overdense fields
\citep{stanway08a}. Although none of the optically identified LBGs
were detected in the 37.7\,GHz band, a faint CO(2-1) emission line was
detected at the same redshift as the optically identified overdensity
(and subsequently reconfirmed in independent data, Stanway et al., in
prep), shown in figure \ref{fig:redxy1040}. With no optical or
near-infrared counterpart in the available imaging and a inferred
molecular hydrogen gas mass of $2\times10^{10}$M$_{\odot}$ this line
emitter is tracing previously undetectable baryonic mass.

The assumption that there is a large reservoir of undetected material
in the structures is consistent with the work of \citet{thomas05},
which, through analysis of fossil populations in nearby galaxes,
determined that massive galaxies found in high density environments
formed the majority of their stars at $z>4$. Further support for such
old star formation in massive galaxies is provided by the distant red
galaxy (DRG) population at $z=2-3$. \citet{labbe05} observed DRGs
which had UV to 8\,$\mu$m spectral energy distributions fitted best by
an old stellar population with very little current star
formation. From the \citet{bruzual03} models used, they infer that the
majority of the stellar mass was in place by $z\sim5$ or earlier
(depending on the metallicity assumed). Similarly, \citet{panter07}
report that the most massive systems seen today formed the majority of
their stars at $z>4$, from analysis of 30,000 galaxy spectra from the
Sloan Digital Sky Survey. \cite{stark07a} also predict that a
significant fraction of the mass at $z\sim5$ is not being observed in
Lyman break galaxy surveys. Assuming that galaxies seen here at
$z\sim5$ are mainly young, the typical stellar mass found by
\citet{verma07} was a few $10^{9}$\,M$_{\odot}$. If this mass is
applied to the galaxies in the redshift overdensities, and it is
assumed that over time they will merge into one of the most massive
systems today (a best-case scenario to accumulate the most mass), the
observed stellar mass at $z=5$ would only account for around 2\% of
the final baryonic mass of a massive galaxy at $z=0$. Although the
structures may extend beyond the 44 arcmin$^2$ field of view of the
imaging or a fraction of the galaxies could be older and more massive
than typical of the population, there is still large fraction of the
mass unaccounted for.

The apparent problem of starburst synchronisation can be avoided if
the galaxies in the overdensities have older stellar populations with
longer-lived starbursts than are typical of the population as a
whole. Although the majority of LBGs at $z\sim5$ studied by
\citet{verma07} are young, a number of older, more massive systems
were identified. Similarly, individual relatively bright objects at
$z\sim6$ have been observed by \citet{yan06} and \citet{eyles05} and
were found to have older and more massive stellar populations. Finding
individual old galaxies suggests that the formation and assembly of
massive galaxies had already started by $z\sim10$. As such sources are
expected to be in the minority of the $z\sim 5$ population,
discovering a connected system of old galaxies would pose significant
problems for our understanding of galaxy and structure formation.

To distinguish between these two scenarios, multiband SED fitting of
the sources is required. With the additional deeper near infrared and
Spitzer IRAC imaging that is currently being added to this extensive
data set, it will be possible to probe the rest-frame UV to optical
spectral energy distributions of these galaxies and determine whether
they have young or old stellar populations.

\section{$z\sim6$ Galaxies}
\label{sec:z6}

As noted previously in section \ref{sec:sample}, the imaging dataset of $V$, $R$, $I$, $z$, $J$ and $K_{s}$-bands can be used to select high redshift objects at $z\sim6$ using an $I_{AB}-z_{AB}$ colour cut. A total of 73 candidates were observed spectroscopically with an $I_{AB}-z_{AB}>1.3$ and non detections in the $V$ and $R$-bands. From this sample two objects were confirmed to be galaxies at $z\sim6$. 

A galaxy with an observed $z_{AB}=25.7$ and a colour of $I_{AB}-z_{AB}=1.4$ was confirmed to be at $z=5.99$ with strong line emission and weak continuum at longer wavelengths but none at shorter wavelengths. With a Ly$\alpha$ flux of $1.9\times10^{-17}$ ergs cm$^{-2}$ s$^{-1}$, the rest-frame equivalent width is 46\AA\ measured using the same method as the $z\sim5$ galaxies (section \ref{sec:ew}). The second galaxy was confirmed with a weaker Ly$\alpha$ line and continuum break at $z=6.41$. This galaxy was observed to have $z_{AB}=24.9$ and a colour of $I_{AB}-z_{AB}>2$. The line flux and rest-frame equivalent width of this emission are $9.4\times10^{-18}$ ergs cm$^{-2}$ s$^{-1}$ and 8\AA \ respectively.

\begin{figure}
\begin{center}
\includegraphics[width=8cm]{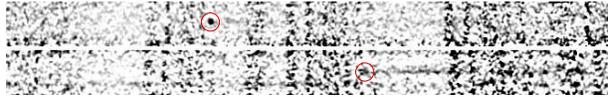}
\caption{2D spectra of the LBGs at $z=5.99$ (top frame) and $z=6.41$ (bottom frame). Both show the spectral region between 7800\AA\ and 9800\AA.}
\label{fig:z6gals}
\end{center}
\end{figure}

Of the other candidates observed, 8 were at low redshift and the
remaining 63 had signal-to-noise too low to be able to identify the
sources. The large number of unidentified objects is unsurprising
given the decreasing response of the FORS2 instrument beyond
9000\AA. With better instrumental responses and a wider range of
wavelengths uncluttered by sky lines at shorter wavelengths, it is
markedly easier to study $z\sim5$ galaxies than their higher redshift
counterparts, even those observed when the Universe was just
200-250\,Myrs older.

\section{Conclusions}
\label{sec:conclusions}

We have carried out a large spectroscopic survey of 10 widely-separate
fields targeting $z\sim5$ LBGs. The candidates were selected using
deep optical imaging combined with a colour selection tuned to
identify the Lyman Break at $z\sim5$. A total of 70 LBGs at
$4.6<z<5.6$ have been identified, 38 galaxies with Ly$\alpha$ emission
ranging in strength from $2.6\times 10^{-18}$ to $7\times 10^{-17}$
erg cm$^{-2}$ s$^{-1}$ (luminosities between $10^{42}$ and $10^{43}$
erg s$^{-1}$) , and 32 were identified by their strong continuum break
at a rest wavelength of 1216\AA. An additional two galaxies were
confirmed to be at $z\sim6$, selected for their red $I-z$ colours. A
comparison of the equivalent width distribution of the Ly$\alpha$
emission of the confirmed $z\sim 5$ LBGs and that of $z\sim 3$ LBGs
from \citet{shapley03} shows that the confirmed higher redshift objects
are twice as likely to have large ($>20$\AA) equivalent width
Ly$\alpha$ emission. Just over half of the $z\sim 5$ objects have
Ly$\alpha$ lines stronger than this. The largest measured equivalent
width is $W_o=122\pm18$\AA. However, including plausible corrections
for the spectroscopically-unconfirmed objects which are nevertheless
at these redshifts removes any significant difference in the fraction
of sources with large ($>20$\AA) equivalent widths in both samples.

A comparison of the average optical spectra and average
$I_{AB}-K_{AB}$ colours of LBGs with and without detectable Ly$\alpha$
emission showed that the break-only galaxies had a redder colour and
hence steeper rest-frame UV continuum slope than the line emitting
galaxies.  It is difficult for stellar age differences alone to
explain this difference given the statistics on the ages of sources
from \citet{verma07}. It can be straightforwardly explained by
variations in reddening (a difference of E(B-V)$\sim0.2$ assuming a
\citet{calzetti00} reddening law can account for the difference in
colours) and/or metallicity between the two subsamples, though age could
clearly also play a part. A similar, but smaller colour difference is
seen between samples of line-emitting and break-only LBGs at lower
redshifts.

If the local correlation between UV-slope and metallicity is assumed
to be applicable at high redshifts, a typical $z\sim 5$ LBG has a
metallicity of approximately 0.1-0.2 of the solar value, approximately
a third that of the metallicity of $z\sim 3$ LBGs.

The morphology of the confirmed $z\sim5$ galaxies was studied using high
resolution HST/ACS images and was found to be similar to that seen in
larger photometric samples with a typical half-light radius of 0.15$''$ or
1\,kpc. However, a higher fraction of multiple and disturbed systems was
observed (16 out of 18 LBGs) suggesting such systems are more likely to
be confirmed spectroscopically, with a greater incidence of Ly$\alpha$
emission. A comparison of the multiple $z\sim5$ systems with local
analogues of LBGs suggests that the components identified in the high
resolution imaging could be either super-starburst regions within a larger
structure with a surface brightness below the detection limit, or part
of merging objects. 

The surface densities of LBGs in two of the ten fields are substantially
higher than in the rest.  In both cases the fields contain a clear
excess of LBGs over a narrow range in redshift ($z=5.05\pm0.1$ and
$z=5.15\pm0.05$). Although these are coherent structures, they may
extend beyond the observed fields (with scale sizes of 20-50 Mpc) and
cannot be bound or virialised systems. Given the distances between the
LBGs within these excesses and the expected short (no more than a few
tens of Myr) timescale for typical $z\sim5$ LBGs, the excess of
systems cannot be caused by gravitational interactions between the
LBGs triggering starbursts. It is more likely that the LBGs are
comparatively rare UV-luminous markers for a much larger mass (both
baryonic and dark) of UV-dark material in these volumes.

In addition to the survey for $z\sim 5$ LBGs, we use the same photometric
data to carry out a comparatively limited search for $z\sim 6$ LBGs by
looking for objects with spectral breaks between the $I$ and $z$-band
images. We confirmed two such galaxies, one at $z=5.99$ with a Ly$\alpha$
line flux of $1.9\times 10^{-17}$ erg cm$^{-2}$ s$^{-1}$ and $z_{AB}=25.7$
and another at $z=6.41$ with a much weaker Ly$\alpha$ line ($9.4 \times
10^{-18} $ erg cm$^{-2}$ s$^{-1}$), but considerably brighter continuum
($z_{AB}=24.9$).

\section*{Acknowledgements}

LSD acknowledges support from P2I. ERS acknowledges support from STFC.
Based on observations made with ESO Telescopes at the La Silla and
Paranal Observatory under programme IDs 166.A-0162 and
175.A-0706. Also based on observations made with the NASA/ESA Hubble
Space Telescope, obtained at the Space Telescope Science Institute,
which is operated by the Association of Universities for Research in
Astronomy, Inc., under NASA contract NAS 5-26555. These observations
are associated with program 9476. We thank the members of the EDisCS
collaboration for creating an imaging dataset with a usefulness far
beyond their original intent. The astronomical table manipulation and
plotting software TopCat \citep{taylor05} was used in the analysis of
this data.  The Dark Cosmology Centre is funded by the Danish National
Research Foundation.

\label{lastpage}

\end{document}